\documentclass[]{article}

\pdfoutput=1
\usepackage{authblk}
\usepackage{fullpage}
\usepackage{amsmath}
\usepackage{booktabs}

\usepackage{caption}
\usepackage{graphicx}

\usepackage[hidelinks]{hyperref}

\usepackage[round, semicolon]{natbib}

\title{Using Convolutional Neural Networks to Develop Starting Models for 2D Full Waveform Inversion}
\author[1]{Joseph P. Vantassel}
\author[1]{Krishna Kumar}
\author[2]{Brady R. Cox}
\affil[1]{The University of Texas at Austin}
\affil[2]{Utah State University}

\begin{document}

\maketitle

\begin{abstract}
Non-invasive subsurface imaging using full waveform inversion (FWI) has the potential to fundamentally change engineering site characterization by enabling the recovery of high resolution 2D/3D maps of subsurface stiffness. Yet, the accuracy of FWI remains quite sensitive to the choice of the initial starting model due to the complexity and non-uniqueness of the inverse problem. In response, we present the novel application of convolutional neural networks (CNNs) to transform an experimental seismic wavefield acquired using a linear array of surface sensors directly into a robust starting model for 2D FWI. We begin by describing three key steps used for developing the CNN, which include: selection of a network architecture, development of a suitable training set, and performance of network training. The ability of the trained CNN to predict a suitable starting model for 2D FWI was compared against other commonly used starting models for a classic near-surface imaging problem; the identification of an undulating, two-layer, soil-bedrock interface. The CNN developed during this study was able to predict complex 2D subsurface images of the testing set directly from their seismic wavefields with an average mean absolute percent error of 6\%. When compared to other common approaches, the CNN approach was able to produce starting models with smaller seismic image and waveform misfits, both before and after FWI. The ability of the CNN to generalize to subsurface models which were dissimilar to the ones upon which it was trained was assessed using a more complex, three-layered model. While the predictive ability of the CNN was slightly reduced, it was still able to achieve seismic image and waveform misfits comparable to the other commonly used starting models. This study demonstrates that CNNs have great potential as a tool for developing good starting models for FWI, which are critical for producing accurate FWI results.
\end{abstract}

\pagebreak

\section{Introduction}
Non-invasive subsurface seismic wave imaging has the potential to revolutionize engineering site characterization and design by providing high-resolution 2D/3D maps of subsurface stiffness. While various techniques exist for subsurface imaging, one of the most promising is full waveform inversion (FWI). FWI can be considered superior to other imaging techniques in two specific ways: first, it uses all of the information available in the seismic wavefield (i.e., phase and amplitude) in contrast to other approaches which only use certain components (e.g., first arrivals) and, second, it produces a map of the site's engineering properties (e.g., stiffness) which can be used directly for subsequent design, in contrast to other approaches which produce only a visual image. FWI can be described as a data-fitting procedure whereby a synthetic seismic wavefield, generated by numerically solving the associated wave equations, is matched to an experimental seismic wavefield acquired in the field. The matching process involves iteratively modifying an assumed starting model through which the synthetic waveforms propagate until the synthetic and experimental wavefields are in acceptable agreement, as determined by the selected wavefield misfit function. Once the iterative optimization process is complete, the final modified model is considered to be an accurate representation of the subsurface. It is important to note that the vast majority of the literature related to FWI has focused on developing and improving the procedures to modify the initial starting model (i.e., improve the waveform matching process), this paper, in contrast, focuses on the selection and importance of the starting model itself.

FWI consists of two stages: (1) propagation of a synthetic seismic wavefield through an assumed subsurface model, and (2) optimization of the assumed model to reduce the misfit between the synthetic and experimental wavefield. Stage 1, more commonly referred to as the forward problem, involves solving the appropriate wave equations using a numerical technique. One of the most common techniques, and the one adopted in this study, is the finite difference method with a staggered-grid discretization \citep{virieux_p-sv_1986, levander_fourth-order_1988}. In practice, various numerical techniques (e.g., finite-element, spectral-element) can be used to solve for the model’s seismic wavefield with acceptable accuracy, thereby making the choice of any particular technique a secondary concern. Stage 2, optimization of the assumed model, may also use a variety of numerical techniques, however, the convergence of these techniques is related to the specific problem under consideration and is not guaranteed generally \citep{nocedal_numerical_2006}. Thus, the method of optimization is of far greater importance to the accuracy of the final solution than the numerical technique used to solve the forward problem. The optimization techniques used for FWI can be grouped into two broad categories of search methods: global and local. Global search methods typically search a large parameter space for the solution with the absolute minimum misfit. Global search methods, while more rigorous than local search methods, come at a significant computational expense, especially when the forward problem is complex and the optimization involves a large number of unknowns, both of which are true for FWI. Yet, despite these complicating factors, some researchers have applied global search techniques to the FWI problem, such as simulated annealing \citep{datta_estimating_2016, tran_two-dimensional_2012}, particle swarm intelligence \citep{mojica_towards_2019}, and genetic algorithms \citep{sajeva_estimation_2016}. However, researchers have more commonly used local search methods which, as their name implies, explore potential solutions located in the vicinity of a starting model in order to minimize the solution’s misfit \citep{nocedal_numerical_2006}. Such techniques are generally faster than their global counterparts, as they typically require solving fewer forward problems, however, without a good starting model these methods are more likely to be trapped in a local minimum that prevents them from converging to the true solution. Thus, a key factor that controls the convergence of the inversion process when using local search algorithms is the selection of an appropriate starting model \citep{shah_quality_2012}. Intuitively, if the starting model is close to the true solution, FWI is more likely to converge to that true solution, whereas, if the starting model is far from the true solution it may become trapped and converge to a local minimum. Therefore, when local search algorithms are employed for FWI, as is most commonly the case, it is of paramount importance to select a starting model that is a reasonable approximation of the true solution.

In addition, it is important to discuss the consequences of the inverse problem's non-uniqueness. The non-uniqueness of the inverse problem, which is a consequence of its ill-posedness \citep{hadamard_lectures_1923}, refers to the possibility that significantly different solutions of the inverse problem (e.g., subsurface seismic images) can rank as equivalent in terms of the inversion's objective function (e.g., seismic wavefield misfit). From a practical perspective, non-uniqueness poses a serious threat to the ability of seismic imaging to produce results that not only fit the measured seismic wavefield (i.e., have low seismic wavefield misfits) but are also consistent with the subsurface structure; the former, while necessary for producing acceptable results, does not necessarily guarantee the latter. While a full discussion regarding the consequences, mitigation, and quantification of non-uniqueness in FWI is much needed, it is unfortunately beyond the scope of this paper. Instead, this brief section is simply to remind the reader of the existence and consequences of non-uniqueness with respect to subsurface seismic imaging.

Selecting an acceptable starting model in FWI is non-trivial as, at best, its fitness may not be apparent until after attempting a time-consuming inversion process and, at worst, its appropriateness (or inappropriateness) may not ever be known. While traditional subsurface characterization techniques (e.g., refraction, first-arrival travel-time tomography, surface wave methods) have been shown to be useful for developing starting models \citep{groos_application_2017, kohn_comparison_2019, pan_high-resolution_2019, wang_tunnel_2019}, they can be time-consuming and require significant expertise to produce acceptable results, thereby making them less than ideal for application in practice, as FWI itself is already rather complex and time-consuming. As an alternative, we propose the use of convolutional neural networks (CNNs) as a tool to quickly develop acceptable starting models for FWI. There are a few properties of CNNs that make them particularly well-suited for such an application. First, CNN’s have shown remarkable performance in the field of image recognition and classification due to their ability to distill (i.e., encode) complex images to a finite number of meaningful representations \citep{chollet_deep_2018}. These meaningful representations can then be combined (i.e., decoded) to create predictions. Second, CNNs developed and trained by researchers with machine learning and subsurface imaging experience can be easily reused by other researchers (with little to no knowledge of machine learning) for their problems. The ability to reuse fully trained CNNs alleviates the problematic expertise requirements imposed by traditional methods. Third, due to the mathematical simplicity of the CNN (i.e., a series of simple tensor operations), starting models for FWI can be produced almost instantaneously for the end user, providing an extremely rapid and convenient alternative to traditional methods. Fourth, due to the modular nature of CNNs, CNNs trained for a particular purpose can readily be expanded by including older networks as components of newer networks (e.g., Simonyan and Zisserman, \citeyear{simonyan_very_2015}) significantly reducing the computational cost of training a new network from scratch, while extending an already successful network to a new purpose \citep{chollet_deep_2018}. For example, a network such as that trained in this study, which focused on two-layered models with an undulating soil-rock interface, can be easily extended to models with multiple interfaces without requiring an entirely new network to be designed and trained. For these reasons, we believe that CNNs show promise as an alternative for developing realistic FWI starting models, and this paper aims to assess the fitness of CNNs towards this end.

To have a basis for comparison, it was desired to compare the novel CNN approach with traditional methods presented in the literature for developing FWI starting models. To do so, 28 works from the literature were identified pertaining to 2D FWI, of which ultimately 12 were considered applicable to the current study, as they specifically addressed near-surface characterization (i.e., depths less than 50 m) using a local search algorithm. These 12 works were grouped further based on their use of either synthetic (where the true model is known) or real experimental data (where the true model is unknown). Works which presented both synthetic and real-world examples were counted twice; once as a member of both groups. Eight of the 12 works contained a synthetic example (Group 1), whereas 10 of the 12 works contained a real example (Group 2). Of the eight works in Group 1: two assumed a constant starting model \citep{kallivokas_site_2013, kucukcoban_full-waveform_2019}, three assumed a starting model increasing linearly with depth \citep{groos_role_2014, tran_sinkhole_2013, tran_site_2012}, and three assumed a smoothed version of the true model \citep{asnaashari_regularized_2013, dokter_full_2017, pan_high-resolution_2019}. Of the 10 works in Group 2: three assumed a constant starting model \citep{kallivokas_site_2013, kucukcoban_full-waveform_2019, tran_characterization_2019}, three assumed a linear trend with depth \citep{dokter_full_2017, tran_sinkhole_2013, tran_site_2012}, and four developed a starting model from other 1D/2D seismic methods \citep{groos_application_2017, kohn_comparison_2019, pan_high-resolution_2019, wang_tunnel_2019}. From this review, the three most commonly used categories of FWI starting models were determined to be constant, linear, and those developed from other 1D/2D seismic methods. Therefore, the starting models developed from the CNN will be compared to starting models derived from each of these three categories.

This study assesses the fitness of using CNNs as a tool for developing starting models for 2D FWI. We begin with the presentation of CNN background information and justification of the selected CNN architecture, followed by the development of the synthetic seismic wavefield-image pairs used for training and validation. The efficacy of the CNN's training is tested by using subsurface models with similar characteristics to those on which it was trained (i.e., two-layered models consisting of soil-over-rock with an undulating interface), although separate from those in the training and validation sets. To assess how the starting model from the CNN compared to other common starting models, two examples: one consisting of a 2-layered system (i.e., similar to the subsurface models on which it was trained), and a second consisting of a 3-layered system (i.e., dissimilar to the subsurface models on which it was trained) are inverted using FWI to directly compare the CNN's prediction with those from three other starting models commonly used in the literature.

\section{Convolutional Neural Networks for Subsurface Seismic Imaging}

Non-invasive seismic testing involves generating stress waves and recording the resulting wavefield on sensors placed at the surface. The process of acquiring the seismic wavefield can be described mathematically with the function $F(m,s)=w$, where $m$, a tensor of subsurface material properties (i.e., the subsurface seismic image), and $s$, a tensor describing the seismic source(s), is transformed into $w$, a tensor of seismic wavefield measurements (i.e., the seismic wavefield). It is important to note that $F$, because of its generality, can and is used in this section to describe both the physical acquisition of data in the field and the numerical solution of the forward problem.

Subsurface seismic imaging seeks to perform $F$ in reverse (i.e., $F^{-1} (w)=m,s$) with the purpose of recovering the subsurface seismic image ($m$) and the signature of the seismic source(s) ($s$) from the recorded seismic wavefield ($w$). However, since it is straightforward to record $s$ in the field, it does not need to be treated as an unknown for the seismic imaging problem; instead we redefine the seismic imaging problem as $I(s,w)=m$, such that $s$ is now treated as known.

In the traditional approach, the solution of the seismic imaging problem ($I(s,w)$) is the seismic image ($m$) which minimizes the error between the measured and predicted seismic wavefields. The error for the traditional imaging problem can be expressed mathematically as:
\begin{equation}
E_{traditional} (m)=f(F(m,s),w)
\end{equation}
where $m$ is the subsurface seismic image being assessed, $f$ is the seismic wavefield error function (e.g., root-mean-square error), $F$, in this context, is the associated numerical forward problem, $s$ is the source description, and $w$ is the recorded seismic wavefield. Note the traditional approach does not attempt to solve the seismic imaging problem explicitly, thereby demanding each new problem, even those which may be quite similar, to be solved independently.

An alternative to the traditional approach is the learning approach (i.e., the one used in this study), where a training set of N ground-truth seismic images ($m_{n=1 to N}$) and their corresponding source descriptions ($s_{n=1 to N}$) and seismic wavefields ($w_{n=1 to N}$) are used to solve for an approximation of $I$ directly. In the learning approach, $I$ is replaced with a deep learning approximation ($R_{\theta}(s,w)  \approx I(s,w)$). Stated another way, the fundamental purpose of the learning approach is to find $R_{\theta}$, which is the best possible replacement for $I$ based on the information provided in the training set. In practice, $R_{\theta}$ is selected such that it minimizes the average error between the ground-truth seismic images and those predicted by the network across the entire training set. The learning error to be minimized to determine $R_{\theta}$ can be expressed mathematically as:
\begin{equation}
E_{learn}(\theta) = \frac{1}{N} \sum_{n=1}^{N}f(m_n,R_{\theta}(s_n,w_n ))    
\end{equation}
where $\theta$ is the set of all weights and biases which define $R_\theta$, $m_n$ is the nth seismic image in the training set, $s_n$ is the nth source description in the training set, $w_n$ is the nth ground-truth seismic wavefield in the training set, and $f$ is the seismic image error function (e.g., mean absolute error). Once the training is complete and $E_{learn}$ has been satisfactorily minimized, the resulting $R_\theta$ is a direct approximation of $I$ and can be used to predict subsurface images directly from seismic measurements. Note that in this study, for simplicity and without loss of generality, we group $s$ and $w$ together into a single observable which we hereafter refer to as the seismic wavefield. The remainder of this work focuses on explaining the development and application of the learning approach to develop starting models for 2D FWI, and comparing the performance of those models to commonly used alternatives.

\section{Architecture of the Selected Convolutional Neural Network}

CNNs are composed of a series of network layers. Each layer in the network performs a differentiable mathematical operation (convolution) on its input to identify/encode specific features. Layers which reside earlier in the network tend to capture less-complex, lower-level features, such as edges, whereas layers located later in the network tend to capture more-complex, higher-level features, such as shapes and objects. Therefore, the ability of a CNN to identify specific features is closely related to its architecture. However, the process of selecting a network architecture is not straightforward, as it relies on the expertise and judgment of the designer(s) as well as trial and error. While some have attempted to fully automate the network selection process, it remains an open and challenging area of research \citep{elsken_neural_2019}. The complexity of the network architecture selection process makes it important to note that for this study many different architectures, some 30 variations in total, were considered prior to final selection. Considered architectures included: those with encoding structures consisting of convolution layers of varying shape, type, and number, and decoding structures consisting of densely connected layers of varying size and number, as well as deconvolutional layers of varying shape and number. Importantly, of all of the network architectures considered, the best performing and, as it happens, one of the simplest is the one presented here.

\begin{figure}[t]
	\includegraphics[width=\textwidth]{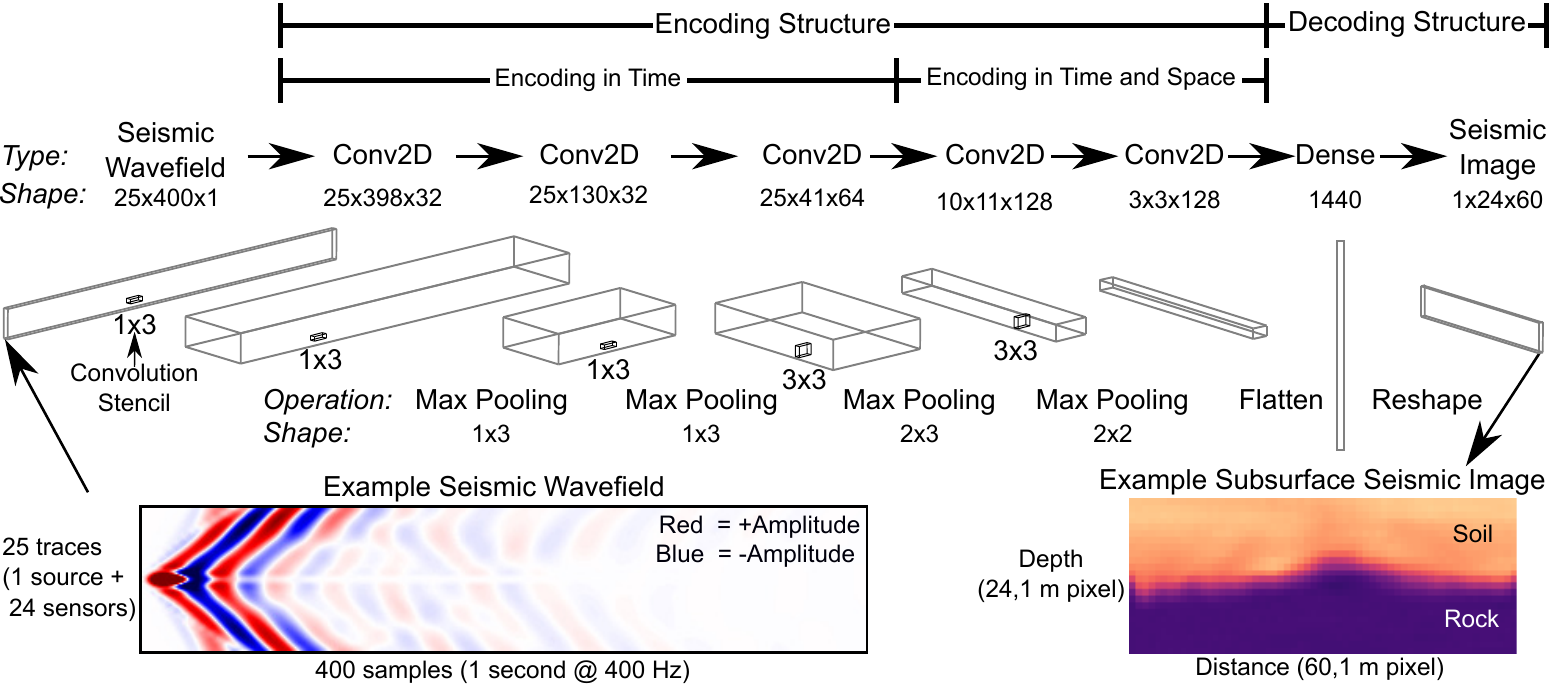}
	\caption{Schematic illustrating the architecture of the convolutional neural network (CNN) used to predict a 2D subsurface seismic image given a seismic wavefield. The predicted subsurface seismic image can be used to create a starting model for 2D full waveform inversion (FWI). The seismic wavefield was selected to be of shape 400x25x1, which physically corresponds to 25 one second recordings sampled at 400 Hz made from one source (located at the model's center) and 24 sensors (spaced at 2 m and located about the model's center). The seismic image was selected to be of shape 1x24x60, which corresponds to a physical model 24-m deep, 60-m wide, and discretized to a 1-m pixel.}
	\label{fig:1}
\end{figure}

The architecture of the network used in this study consists of an encoding structure composed of five convolutional layers interleaved with four max pooling layers followed by a decoding structure composed of a single densely connected layer. A schematic of the network is shown in Figure \ref{fig:1} and a more detailed summary is provided in Table \ref{table:1}. The input of the CNN is a 3-dimensional (3D) tensor containing the seismic wavefield recorded by a standard linear array of 24 surface sensors. As this study was focused primarily on examining the feasibility of using CNNs as a tool to develop starting models for 2D FWI, we only consider a seismic wavefield input of a single shape (25x400x1). In physical terms, these dimensions represent 25 one-second long time histories sampled at 400 Hz from a single source recorded by 24 sensors. For this study, the 24 sensors were placed at 2 m intervals and located about the model's center. An example synthetic seismic wavefield for a source excited at the middle of the linear sensor array (i.e., at the center of the model) is shown in the lower left of Figure \ref{fig:1}. An important and novel feature of the CNN architecture, is the use of atypical shapes for the first three convolutional and first two max pooling layers. These atypical shapes of 1x3 were selected over traditional square shapes (e.g., 3x3) to ensure that a single time record, which is likely over-sampled in time, is sufficiently simplified into a feature-like representation prior to its convolution with other adjacent time records. Stated another way, the selected architecture seeks to simplify each time series independently to allow the network to learn a meaningful time-wise representation prior to learning the relationship between adjacent time series (i.e., space-wise representation). We believe this approach simplifies the complex problem of understanding the meaning of the seismic wavefield into two simpler problems. Furthermore, the division of representation allows the network to remain relatively light and easy to train.

\begin{table}[]
	\centering
	\caption{Detailed description of each layer of the convolutional neural network (CNN), including network layer type and the shape of the input and output tensors.}
	\begin{tabular}{@{}ccc@{}}
		\toprule
		Network Layer Type & Input Shape                                                               & Output Shape                                                                 \\ \midrule
		2D Convolution     & \begin{tabular}[c]{@{}c@{}}25x400x1   \\ (Seismic Wavefield)\end{tabular} & 25x398x32                                                                    \\
		2D Max Pooling     & 25x398x32                                                                 & 25x132x32                                                                    \\
		2D Convolution     & 25x132x32                                                                 & 25x130x32                                                                    \\
		2D Max Pooling     & 25x130x32                                                                 & 25x43x32                                                                     \\
		2D Convolution     & 25x43x32                                                                  & 25x41x64                                                                     \\
		2D Max Pooling     & 25x41x64                                                                  & 12x13x64                                                                     \\
		2D Convolution     & 12x13x64                                                                  & 10x11x128                                                                    \\
		2D Max Pooling     & 10x11x128                                                                 & 5x5x128                                                                      \\
		2D Convolution     & 5x5x128                                                                   & 3x3x128                                                                      \\
		Flatten            & 3x3x128                                                                   & 1152                                                                         \\
		Dense              & 1152                                                                      & 1440                                                                         \\
		Reshape            & 1440                                                                      & \begin{tabular}[c]{@{}c@{}}1x24x60\\ (Subsurface Seismic Image)\end{tabular} \\ \bottomrule
	\end{tabular}
	\label{table:1}	
\end{table}

As is common practice, the size of the convolutional layers which reside deeper in the network increase (i.e., from a typical size of 32 in the first layer to a robust 128 in the final layer) to provide space for the network to store the learned features of the seismic wavefield. Note that the size of a convolutional layer can easily be determined by examining the final dimension of its output shape (e.g., the first convolution layer which has an output shape of 25x398x32, refer to Figure \ref{fig:1} and Table \ref{table:1}, has a size of 32). In addition, it is important to note for those unfamiliar with CNNs that the size and shape of the network's interior layers do not have any physical meaning, but have been selected solely to optimize the network's performance. To transform the learned features of the seismic wavefield (i.e., the encoded representation) into a 2D seismic image (i.e., the decoded representation), the features are flattened from a 3D tensor to a 1D tensor (i.e., vector) and are passed through the decoding structure composed of a single densely connected layer. The densely connected layer contains 1440 nodes, one per seismic image pixel, such that the result can be reorganized into the shape of the predicted seismic image (i.e., 1x24x60). The physical meaning behind the shape of the resulting subsurface seismic images is a model 24-m deep, a typical depth for near-surface characterization efforts (e.g., Tran and McVay \citeyear{tran_site_2012}), 60-m wide, to accommodate the array, source locations, and boundary conditions (discussed later), and discretized with a 1-m pixel size, to ensure a reasonably detailed subsurface image. Note that while a subsurface seismic image of 24 m x 60 m was used in this study, other sizes could have been accommodated by slightly modifying the selected network architecture, for example, by increasing or decreasing the size of the final densely connected layer. An example subsurface seismic image predicted by the CNN from the example seismic wavefield is presented in the lower right corner of Figure \ref{fig:1}.  In order for the CNN to be able to learn how to predict a seismic image (output) from a provided seismic wavefield (input), the network must be trained using input-output pairs referred to as the training set. The details regarding the development of the input-output pairs used for training are discussed in the following section.

\section{Development of Seismic Wavefield-Image Pairs}

\subsection{Seismic Image Development}

\begin{figure}[t]
	\includegraphics[width=\textwidth]{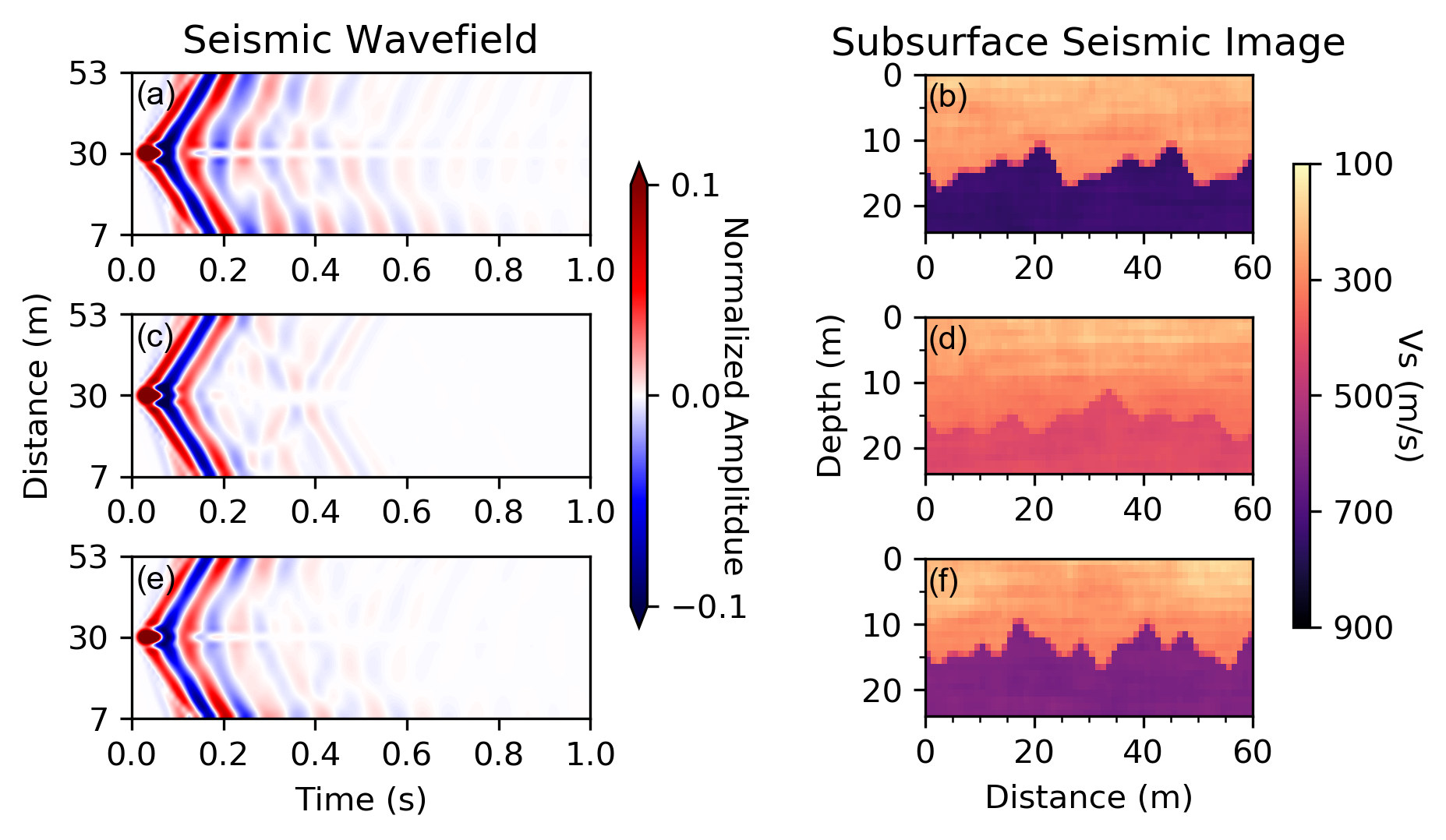}
	\caption{Three example seismic wavefield-image pairs from the soil-over-rock convolutional neural network (CNN) training set. Panels (a), (c), and (e) show the seismic wavefields recorded on a 46-m long array (24 receivers with 2 m spacing) placed about the model's center (spanning from 7 to 53 m). The seismic source placed at the center of the linear array (distance of 30 m) is also included in the seismic wavefield. Panels (b), (d), and (f) are the shear wave velocity (Vs) subsurface seismic images that correspond to the seismic wavefields shown in panels (a), (c), and (e), respectively.}
	\label{fig:2}
\end{figure}

The 100,000 unique seismic models developed in this study to train the CNN are representative of stiff soil over undulating weathered bedrock. The training models were developed automatically using an algorithm described below. Each subsurface model was developed by first building the shear wave velocity (Vs) seismic image then relating compression wave velocity (Vp) and mass density to Vs using simple, common relationships. It is important to clarify here that that the authors, in an effort to avoid confusion, have attempted to use the term image whenever referring to a single subsurface parameter (e.g., Vs), and the term model when referring to the entire subsurface model which, in the isotropic-elastic case used here, requires the definition of Vs, Vp, and mass density. The Vs images were constructed by first assuming the Vs of the soil layer (i.e., the upper part of the model) follows the approximate relationship between Vs and mean effective stress for dense sand \citep{menq_dynamic_2003}. To avoid unrealistically low velocities, the Vs relationship was truncated near the ground surface (i.e., at low mean effective stresses) to ensure no Vs less than 200 m/s. To model a broader range of realistic dense soil-velocities, the approximate relationship was scaled by a random variable, the soil velocity factor, between 0.9 and 1.1. The interface between the upper and lower layers (i.e., soil and weathered rock) was defined using four variables: the average depth to bedrock, which ranged from 5 m to 15 m, and three independent undulation ``frequencies'', which ranged from 1/5 m$^{-1}$ to 1/60 m$^{-1}$. The weathered rock layer (i.e., the lower part of the model) was constrained to have a constant Vs between 360 m/s and 760 m/s. Note the transition between dense-sand and weathered rock represents a considerable impedance contrast (i.e., generally greater than 2). The presence of such contrasts are ubiquitous in near-surface characterization and yet have received limited attention in the literature, making them particularly important to consider here. In summary, the following parameters were varied to define the Vs seismic images: soil velocity factor, average depth to bedrock, bedrock undulation (defined using three independent ``frequencies''), and weathered rock velocity. All of these variables were simulated by randomly drawing from different uniform distributions defined by their upper and lower limits. These variables and their upper and lower limits are summarized in Table \ref{table:2}. To introduce additional complexity to the Vs image, a unique laterally correlated perturbation was applied. The use of the perturbation introduced small-scale irregularities ($\approx$1 -- 2 m in the vertical and $\approx$4 -- 6 m in the horizontal directions) representing stiff and soft inclusions, thereby making these models, in the opinion of the authors, more realistic than those which are typically used in synthetic studies. A sample of three Vs subsurface seismic images created using this methodology and used as part of the training set are shown in panels (b), (d), and (f) of Figure \ref{fig:2}. As mentioned previously, the Vp and mass density images were developed by assuming simple relationships between Vs and the parameter of interest. In particular, the Vp image was developed by assuming a Poisson’s ratio of 0.33 for the soil (i.e., Vs $<$ 300 m/s) and 0.2 for weathered rock (i.e., Vs $>$ 360 m/s). For intermediate materials, the Poisson's ratio was linearly interpolated. The mass density image was developed by assuming a value of 2000 kg/m$^3$ for soil and 2100 kg/m$^3$ for weathered rock. A total of 100,000 unique subsurface models were generated using this approach for the CNN's training and validation, with an additional 100 being generated and set aside for testing.

\begin{table}[]
	\centering
	\caption{Material parameters, along with their upper and lower limits, which were simultaneously varied following their respective uniform distributions to produce 100,000 unique seismic images consisting of stiff soil over undulating weathered bedrock for use as the convolutional neural network's training set.}
	\begin{tabular}{@{}lcc@{}}
		\toprule
		\multicolumn{1}{c}{Material Parameter}                                                                               & Lower Limit                                                                             & Upper Limit                                                                            \\ \midrule
		Soil Velocity Factor (\#)$^a$                                                                                        & 0.9                                                                                     & 1.1                                                                                    \\
		Average Depth to Weathered Rock (m)                                                                                  & 5                                                                                       & 15                                                                                     \\
		Frequency of Weathered Rock Undulation (1/m)                                                                         & 1/60, 1/20, 1/10                                                                        & 1/20, 1/10, 1/5                                                                        \\
		Shear Wave Velocity of Weathered Rock (m/s)                                                                          & 360                                                                                     & 760                                                                                    \\
		\midrule
		\multicolumn{3}{l}{\begin{tabular}[c]{@{}l@{}}$^a$The soil velocity factor was applied to the approximate relationship between mean effective stress\\ and shear wave velocity from Menq (\citeyear{menq_dynamic_2003}) to obtain shear wave velocities for the stiff-soil layer\\ representative of dense sand.\end{tabular}} \\ \bottomrule
	\end{tabular}
	\label{table:2}
\end{table}

\subsection{Seismic Wavefield Development}

To develop a seismic wavefield representative of each subsurface model, a 2D finite difference wave propagation simulation was performed using the open-source software DENISE \citep{kohn_time_2011, kohn_influence_2012}. As mentioned previously, to simplify this initial study we utilized only a single source location (at the center of the model) and a single source wavelet (a 15 Hz high-cut filtered spike). The resulting waveforms were recorded on a 24-channel array of surface sensors with 2 m spacing, centered on the source location. The simulation utilized a 5E-5 second time step for a total duration of 1 second. The simulation used a 6th order finite difference operator in space and a 2nd order finite difference operator in time. To ensure the simulation's numerical stability, each seismic image was resampled from a 1-m pixel down to a 0.2-m pixel. The top of the domain utilized the free-surface boundary condition \citep{levander_fourth-order_1988} and the sides and bottom of the domain were truncated using the perfectly matched layer boundary condition \citep{komatitsch_unsplit_2007}. To reduce the data storage requirements, the time series were down-sampled from that required by the stability of the finite difference simulation (20,000 Hz) to a more manageable and realistic 400 Hz. Three example seismic wavefields are shown in panels (a), (c), and (e) of Figure \ref{fig:2} corresponding to the three Vs seismic images in panels (b), (d), and (f), respectively.

\section{Training of the CNN}

The open-source machine learning library Keras \citep{chollet_keras_2015} was used to construct and train the CNN. As the construction of a CNN using Keras is rather straightforward and the architecture of the model has already been discussed at length, we will focus primarily on the data preprocessing, data augmentation, and network training in this section. First, in preparation for training, as is common practice, the seismic wavefield and Vs seismic images were normalized using a linear transformation to values between 0 and 1 to assist the training process. Note that since this transformation is linear and stored for later reference it can be applied in the reverse direction (i.e., back to physical values) without any loss of generality. Second, the 100,000 original normalized seismic wavefield-image pairs were used to generate an additional 100,000 normalized seismic wavefield-image pairs via a data augmentation procedure, resulting in a total of 200,000 wavefield-image pairs that could be used for training. The data augmentation procedure operated on each of the original wavefield-image pairs, whereby the wavefield and image were both mirrored about their centerlines to produce a new seismic wavefield-image pair for the training set. The data augmentation process can be thought of in physical terms as first viewing a 2D seismic image, like the example shown in Figure \ref{fig:1}, from the original perspective, and then viewing the exact same 2D seismic image from the opposite perspective of stepping through the page, turning around to face the 2D image, and looking at it again from the backside of the page (i.e., the augmented perspective). What was originally on your left is now on your right. In both cases, regardless of the observer's perspective, the underlying subsurface structure and wavefield is unchanged, however, as the CNN is unaware of a problem's underlying physics, data augmentation can be used to increase the size of the training set and increase the CNN's ability to generalize. Using this data augmentation technique, the training data was doubled at minimal computational cost, as it required no additional forward problems to be solved. Third, the 200,000 training pairs were shuffled to ensure the original (i.e., the first 100,000) and the augmented (i.e., the second 100,000) pairs were appropriately mixed. The network was trained using a P100 compute node of Maverick2, a specialized computing resource at the Texas Advanced Computing Center (TACC) designed specifically for machine learning research. The Maverick2 node includes two NVIDIA Tesla P100 GPUs which were employed in parallel to expedite the training process. In total, it took approximately 1 -- 2 hours to train each of the 30 network configurations considered during architecture development. Due to memory constraints, the network training was split into two stages. Stage 1 utilized the first 100,000 shuffled training pairs and Stage 2 the second 100,000 shuffled training pairs. Just as with the network architectures, the training hyperparameters (e.g., number of epochs, optimizer, learning rate, loss function, etc.) were systematically varied to improve performance. Between 10 and 20 different training hyperparameter configurations were considered for each of three finalist network architectures. For brevity, we will only detail the final hyperparameter configuration, which was used for both Stages 1 and 2. The final configuration utilized the Adam optimizer \citep{kingma_adam_2014} with a learning rate of 0.0005. The error function selected was the mean absolute error (MAE), which is the mean of the absolute pixel-by-pixel difference between the true and predicted image. Training included 20 epochs (i.e., complete iterations over the training data) with a batch size of 16. To allow for the identification and mitigation of overfitting throughout the training process, 10\% of the training data from each stage (20,000 models in total) was set aside as a validation set. At the conclusion of training, the network achieved an overall average MAE of 0.0294 on the training data and a MAE of 0.0299 on the validation data, indicating no excessive overfitting and an excellent performance by the CNN to predict the true Vs image even before starting FWI.

\section{Testing the CNN}

\begin{figure}[t]
	\includegraphics[width=\textwidth]{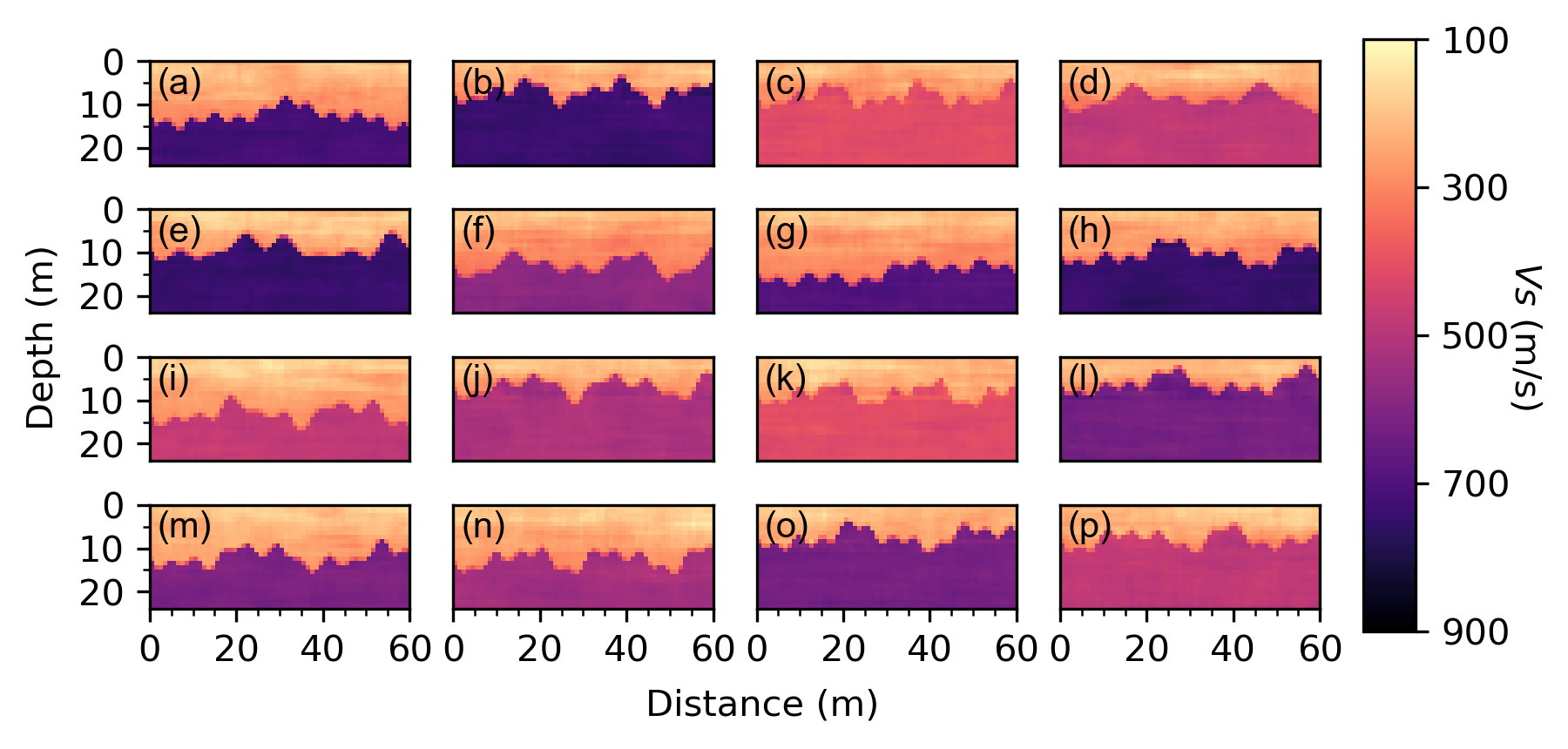}
	\caption{Sixteen randomly selected shear wave velocity (Vs) subsurface seismic images from the convolutional neural network's (CNN's) 100-member testing set.}
	\label{fig:3}
\end{figure}

\begin{figure}[!]
	\includegraphics[width=\textwidth]{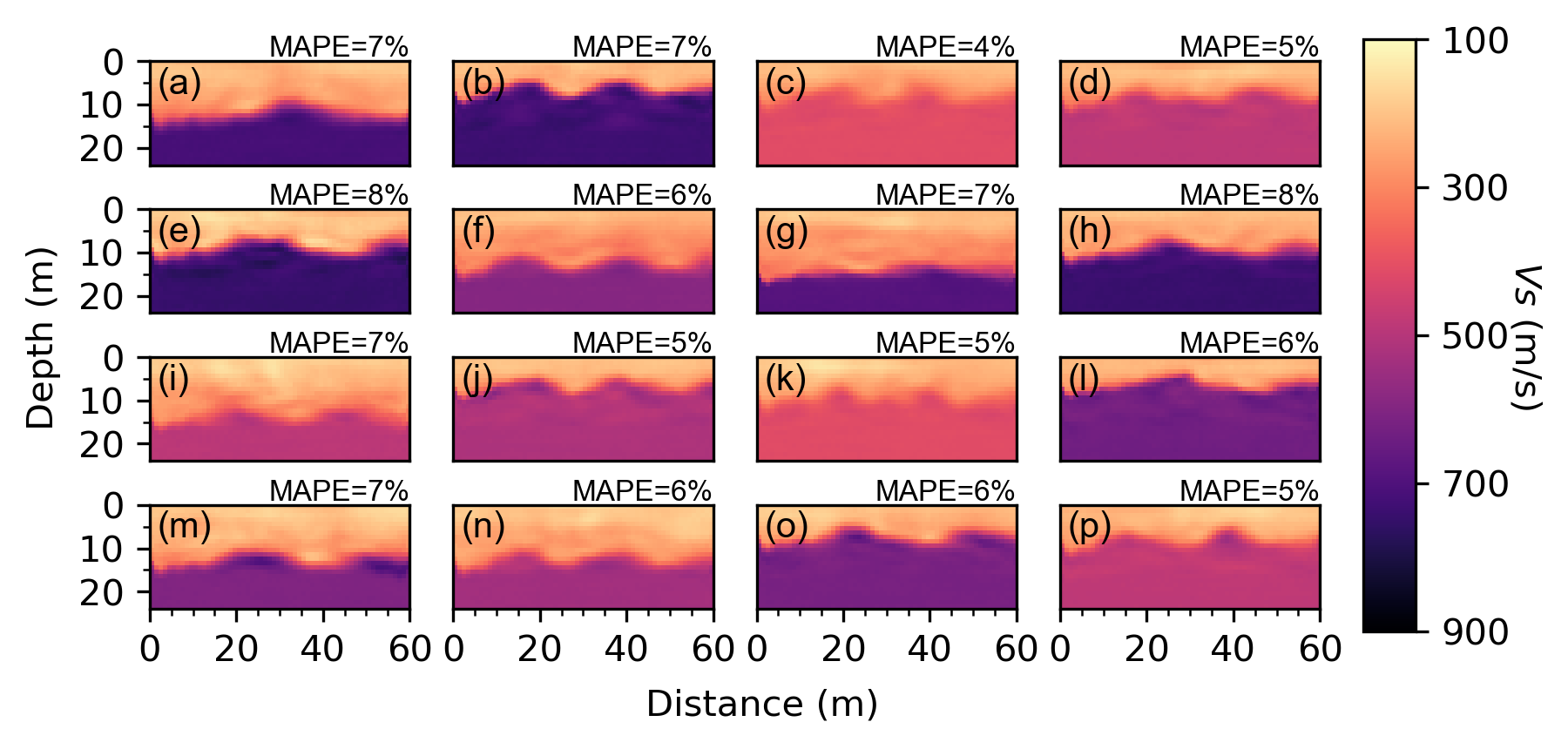}
	\caption{The convolutional neural network's (CNN's) prediction of the 16 true shear wave velocity (Vs) subsurface seismic images presented previously in Figure \ref{fig:3}. Note these predictions are made by the CNN using only the measured seismic wavefield. The mean absolute percent error (MAPE) of each image is presented in the upper right of each panel.}
	\label{fig:4}
\end{figure}

\begin{figure}[t]
	\includegraphics[width=\textwidth]{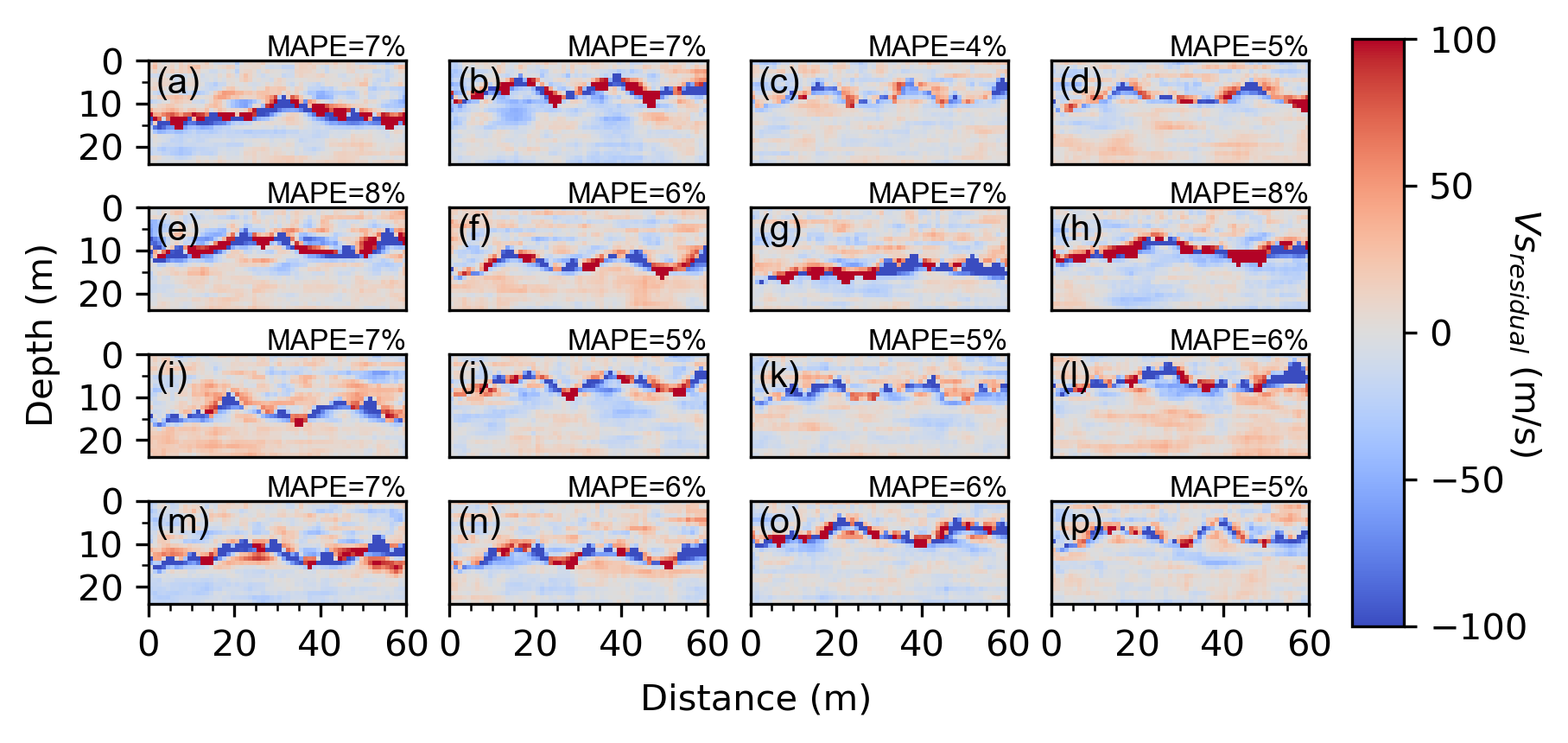}
	\caption{The residual shear wave velocity (Vs\textsubscript{residual}) [i.e., the pixel-by-pixel difference between the convolutional neural network (CNN) predicted Vs image (Vs\textsubscript{predicted}) and the true Vs image (Vs\textsubscript{true})] for the 16 examples from the testing set presented in Figures \ref{fig:3} and \ref{fig:4}. The mean absolute percent error (MAPE) of each image is presented in the upper right of each panel.}
	\label{fig:5}
\end{figure}

\begin{figure}[!]
	\includegraphics[width=\textwidth]{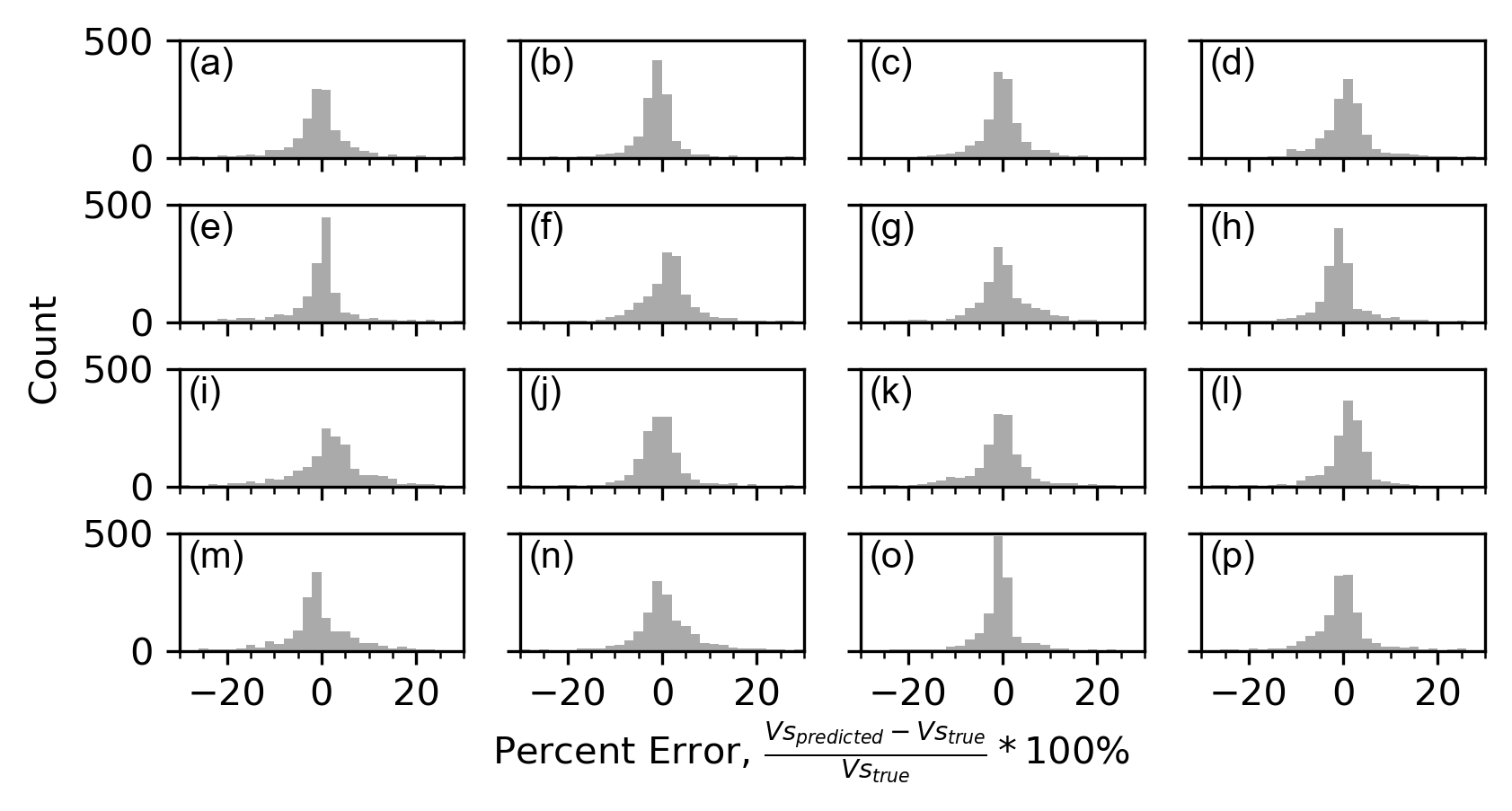}
	\caption{Distributions of the pixel-by-pixel percent error for the 16 examples presented in Figures \ref{fig:3} and \ref{fig:4}.}
	\label{fig:6}
\end{figure}

After training, the CNN was tested using the set of 100 previously developed but unused seismic wavefield-image pairs. Since these Vs images were not used in the testing phase (neither as part of the training set nor as part of the validation set) they provide a more objective assessment of the network's performance than the results presented from the training and validation reported previously. Evaluation of the CNN on the testing data confirmed the CNN's remarkable performance with an overall average MAE of 0.0301, indicating that the validation set used to monitor overfitting in the training stage produced a reasonably objective assessment of the CNN's performance and was not overly affected by information leakage \citep{chollet_deep_2018} caused through architecture selection and hyperparameter tuning. To demonstrate the effectiveness of the CNN in a more visual manner, compare Figure \ref{fig:3}, which contains 16 Vs images from the testing set, with those in Figure \ref{fig:4}, which contains the corresponding 16 Vs images predicted by the CNN. A visual comparison of the two figures shows remarkable similarities between the true and predicted images, with the soil and rock velocities being reasonably resolved in all cases. To provide a quantitative assessment of the CNN's performance in physical units, the mean absolute percent error (MAPE), which is the mean of the absolute value of the pixel-by-pixel percent error, of each predicted image is listed in the upper right of each panel. Note that we report MAPE here rather than MAE, as the MAE used during training has no direct meaning in physical space (only in the normalized space), and therefore is not as useful for comparison of the seismic images in physical units. The values of the MAPE in Figure \ref{fig:4} ranges between 4\% and 8\% (the average across all 100 test images was 6\%) indicating that all of the Vs images predicted by the CNN are a good approximation of the true Vs image, even before FWI. The main distinction between the seismic images in Figures \ref{fig:3} and \ref{fig:4} is at the soil-rock boundary, where the CNN tends to produce a smoothed version of the true rock undulations. To compare the similarities and differences between the true and CNN predicted images more quantitatively, Figure \ref{fig:5} presents the pixel-by-pixel residual Vs (Vs\textsubscript{residual}), which is the difference between the Vs predicted by the CNN (Vs\textsubscript{predicted}) and the Vs from the true model (Vs\textsubscript{true}) (i.e., Vs\textsubscript{residual} = Vs\textsubscript{predicted} - Vs\textsubscript{true}). Note for comparison, the MAPE is again presented in the upper right of each panel. Figure \ref{fig:5} quantitatively confirms the previous qualitative observations that the CNN well represents the soil and rock velocities, however, struggles to exactly resolve the finer undulations in soil-rock interface.  While the MAPE values indicate excellent performance by the CNN, it is also important to examine the distribution of error for potential bias. Figure \ref{fig:6} presents the distribution of the pixel-by-pixel percent error for the same 16 models presented in Figure \ref{fig:3}. The distributions show that the errors in the CNN's predictions are roughly symmetric (i.e., contain equal parts over- and under-prediction) and unbiased (i.e., centered on zero). The authors believe the inability of the CNN to exactly resolve the small-scale features of the true models, particularly at the soil-rock interface, is due at least in part to the physical limitations of the problem as posed, rather than a limitation of the CNN approach itself. In particular, the authors recognize the use of only a single source location with a relatively low frequency (long wavelength) wavelet as the key factor preventing the resolution of small-scale structure. However, despite these limitations, the CNN is shown to perform remarkably well and thereby demonstrate that CNNs are a promising tool for developing starting models for FWI.

\clearpage

\section{Comparison of the CNN-Derived Starting Model with other Starting Models in FWI: Two-Layered Example}

\begin{figure}[!t]
	\centering
	\includegraphics[width=0.9\textwidth]{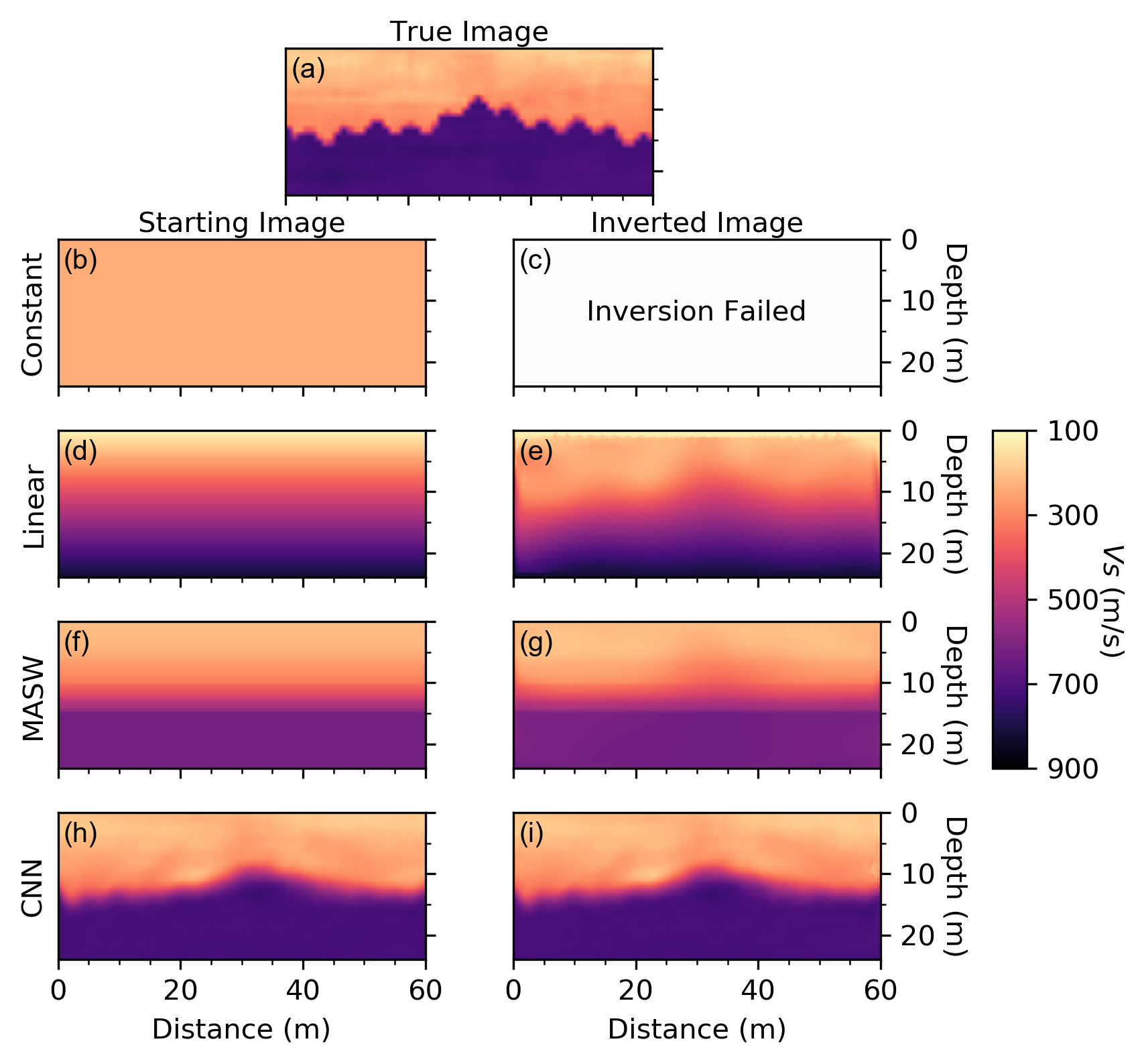}
	\caption{Performance of the convolutional neural network (CNN) in comparison to other typical starting models for a \textbf{two-layered example} from the testing set.  Panels (b), (d), (f), and (h) show the constant, linear, multichannel analysis of surface waves (MASW), and convolutional neural network (CNN) shear wave velocity (Vs) starting image, adjacent to their inverted Vs images after full waveform inversion (FWI) (i.e., panels (c), (e), (g), and (i), respectively). For reference the true Vs image is shown in panel (a). Note that no inverted image is available for the constant starting model, panel (c), as the inversion failed.}
	\label{fig:7}
\end{figure}

To assess the performance of the CNN, we use its predicted Vs image as a starting model for FWI and compare the FWI results with those obtained from three other starting model alternatives previously identified as being commonly used in the literature. The example used for this comparison is the image shown in Figure \ref{fig:3}a, whose CNN prediction in Figure \ref{fig:4}a is shown to have an MAPE = 7\%. To facilitate an easy comparison, the true image is repeated in Figure \ref{fig:7}a and the CNN predicted image in Figure \ref{fig:7}h. The CNN's prediction captures the generally concave down shape of the soil-rock interface and the soil and rock velocities, however, it is unable to accurately resolve the small-scale undulations of the soil-rock interface. The CNN's prediction is compared to three commonly-used 1D starting model alternatives in Figures \ref{fig:7}b, \ref{fig:7}d, and \ref{fig:7}f: constant, linear, and one derived from the multichannel analysis of surface waves (MASW), respectively. The constant starting model shown in Figure \ref{fig:7}b was developed by finding the value of Vs with the least square difference with the true model over the top 10 m (i.e., the predominantly soil-like portion of the model) and then relating the other model parameters (Vp and mass density) to Vs using the simple expressions used to develop the CNN's training set. While this procedure cannot be done in practice, as the true Vs is never known, we did so here to give the constant model the best possible chance of success without requiring the time-consuming trial and error process, as is typically necessary in practice. The linear starting model in Figure \ref{fig:7}d was created in a similar manner to that of the constant model, except the linear trend of Vs having the least square difference with the true Vs was selected and there was no need to truncate the comparison to the top 10 m. Again, this was done to give the linear model the most likely chance of success. The MASW starting model presented in Figure \ref{fig:7}f was developed by first simulating seismic wavefields using four additional source locations (located at 1 m and 5 m off either end of the array) and recording the wavefield on the same 24 receiver array with 2 m spacing. To develop a 1D Vs profile representative of the entire model, the dispersion data from each of the four source locations were processed into a single measure of the site's dispersion data, including estimates of uncertainty, using the open-source Python package \textit{swprocess} \citep{vantassel_jpvantasselswprocess_2021}. The site's dispersion data was then inverted using 3 Layering by Number (LN) parameterizations (LNs of 5, 7, and 9) with three trial inversions per parameterization, following the recommendations of Vantassel and Cox (\citeyear{vantassel_swinvert_2021}). The inversion computations were performed with the Dinver module \citep{wathelet_surface-wave_2004} of the open-source software Geopsy \citep{wathelet_geopsy_2020}. The 100 lowest misfit 1D Vs profiles from each of the three different parameterizations (across all trials) were combined into a single discretized median profile. This 1D discretized profile was then extended over the entire width of the domain to produce the starting model shown in Figure \ref{fig:7}f. To complete the MASW starting model, Vp and mass density were assigned to the Vs image using the same simple rules discussed previously.

To provide a fair comparison between the FWI results obtained from the four starting models, each utilized the same FWI settings and multi-scale inversion workflow. The FWI analysis was performed using the same open-source software DENISE \citep{kohn_time_2011, kohn_influence_2012} as was used for the solution of the forward problem in the development of the training set. To simulate a more realistic real-world application of FWI, the single source applied at the center of the model used during the construction of the CNN testing set was replaced with 25 source locations spaced at 2 m (i.e., 1 m off both ends and between each receiver). The same 15 Hz high-cut-filtered spike wavelet as was used during the construction of the CNN testing set was again used as the source signal. The settings related to the finite difference simulation to solve the forward problem were the same as those discussed with respect to the development of the training set. The optimization algorithm used was the Limited-Memory Broyden-Fletcher-Goldfarb-Shanno (LBFGS) \citep{byrd_limited_1995, broyden_convergence_1970, fletcher_new_1970, goldfarb_family_1970, shanno_conditioning_1970}, where the beta coefficient has been defined using the approach proposed by Polak and Ribiere (\citeyear{polak_note_1969}). The algorithm was permitted to retain (i.e., “remember”) up to 10 updates during LBFGS optimization. To reduce large gradients at the model's source locations, circular preconditioning tapers with a radius of 1 m were applied around each source location. To still permit the model to update in these locations during the inversion, the 25 sources were simulated separately and the preconditioning only applied to the currently active source location. To limit the introduction of spurious small-scale features, the model gradient was smoothed using a 2D Gaussian stencil extending over a 1 m x 1 m area (i.e., recall for solving the forward problem a 0.2-m pixel size was necessary for stability). Inversions utilized a multi-scale workflow \citep{bunks_multiscale_1995} to mitigate the potential for cycle skipping and to improve convergence. The multi-scale inversion workflow used three stages with the overlapping frequency ranges of 0 - 7 Hz, 0 - 13 Hz, and full signal. Each stage was required to consist of at least 10 optimization iterations and was able to conclude after either the change in the misfit became less than 1\% or 100 iterations were performed. The only exception to these rules is if a simulation became unstable in the middle of the stage without yet reaching one of the aforementioned stopping criteria, then the inversion was restarted in the subsequent stage. During the inversion process, all three model parameters (Vs, Vp, and mass density) were allowed to be updated to ensure no penalty was applied to models which started further from the true solution and which therefore likely contained worse estimates of Vp and mass density. As the wavefield is not strongly sensitive to mass density \citep{kohn_influence_2012}, the gradient when applying the mass density update was scaled by a factor of $\frac{1}{2}$ to discourage the creation of artifacts in the density model related to large updates early in the inversion process.

Returning to Figure \ref{fig:7}, the inverted Vs images after the conclusion of the final stage of FWI for each of the four starting models are shown in panels (c), (e), (g), and (i). The inversion of the constant starting model failed during the first iteration of the first stage and no results are available. Several attempts were made to tweak the optimization algorithm to produce a result, however, no solution was possible. We believe that the failure of the constant starting model was related to its significant difference from the true model, despite the efforts discussed previously to ensure the model was the closest possible approximation of the true model's upper 10 m. Inversion of the linear model was successful and results are shown in panel (e). FWI improves the Vs image in two noticeable ways: first, the velocity of the soil material (i.e., approximately the upper 10 m), with the notable exception of the top 1 m, is shown to more closely reflect the true soil velocity and its spatial variability. Second, a vague outline of the soil-rock contrast can be approximately identified. Importantly, however, the velocity of the soil and rock are not well captured and the sharp velocity contrast is poorly resolved. The FWI results obtained from the MASW and CNN starting models, shown in panels (g) and (i), respectively, show very minor updates after inversion. The main change for the MASW model is at the center of the model between 20 and 40 m, where the faint outline of the true model's concave down bedrock structure is visible. However, neither a sharp boundary contrast nor the correct rock velocity is resolved after FWI using the MASW starting model. For the CNN starting model, some minor changes are visible after FWI, for example in the model's upper right corner, however, the model remained largely unchanged through the inversion process.

\begin{figure}[!t]
	\centering
	\includegraphics[width=0.9\textwidth]{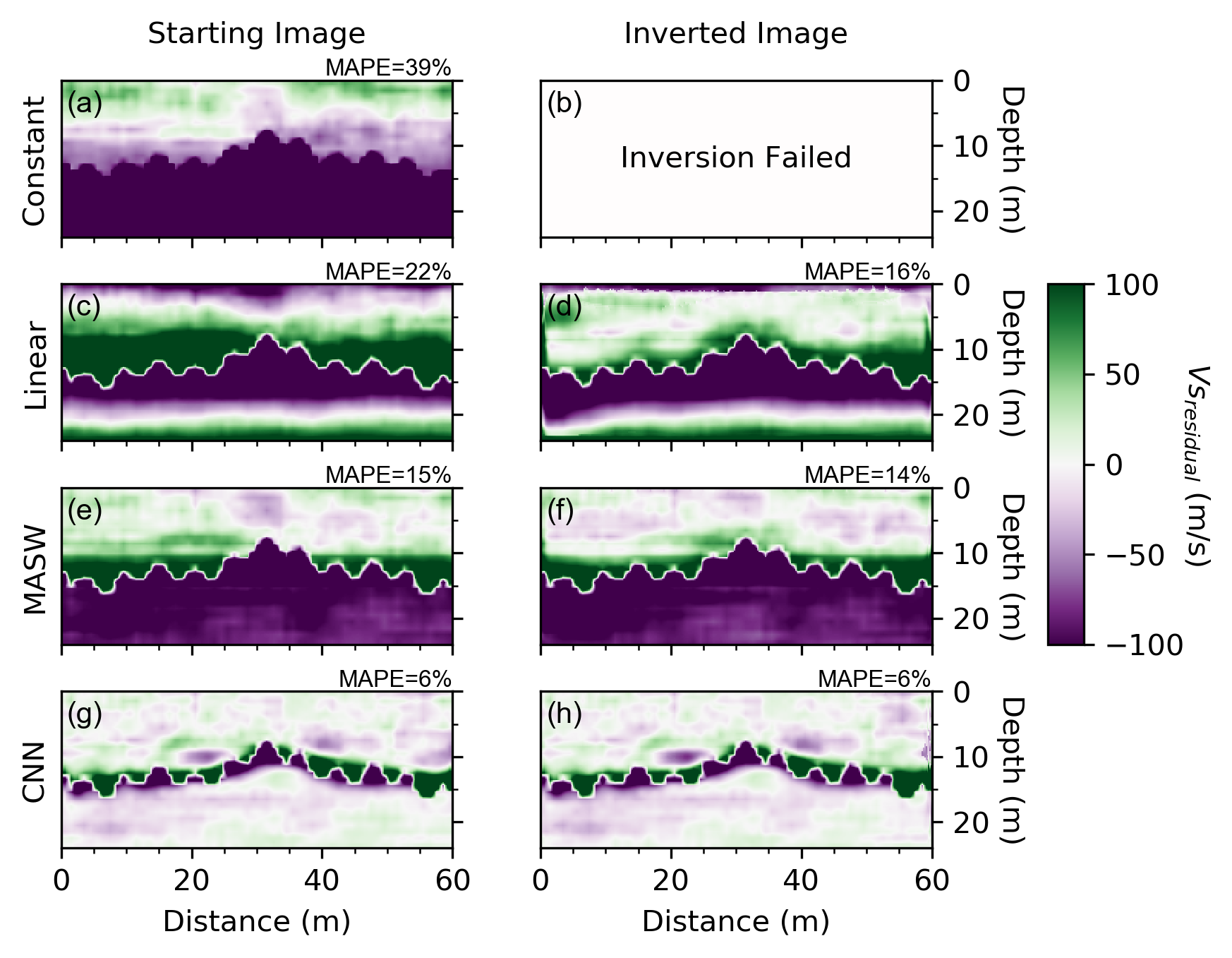}
	\caption{The residual shear wave velocity (Vs\textsubscript{residual}) for the four starting models before and after full waveform inversion (FWI) for a \textbf{two-layered example} from the testing set. Panels (a), (c), (e), and (g) show Vs\textsubscript{residual} for the constant, linear, multichannel analysis of surface waves (MASW), and convolutional neural network (CNN)  starting images, adjacent to that of the inverted images after FWI (i.e., panels (b), (d), (f), and (h), respectively). The mean absolute percent error (MAPE) of each image is presented in the upper right of each panel. Note that the thin region of Vs\textsubscript{residual}=0 which appears to delineate the bedrock interface is purely an artifact of how Vs\textsubscript{residual} is calculated and does not indicate the successful resolution of the boundary. Note that Vs\textsubscript{residual} could not be calculated for the constant starting model, panel (b), as the inversion failed.}
	\label{fig:8}
\end{figure}

\begin{figure}[!t]
	\centering
	\includegraphics[width=0.9\textwidth]{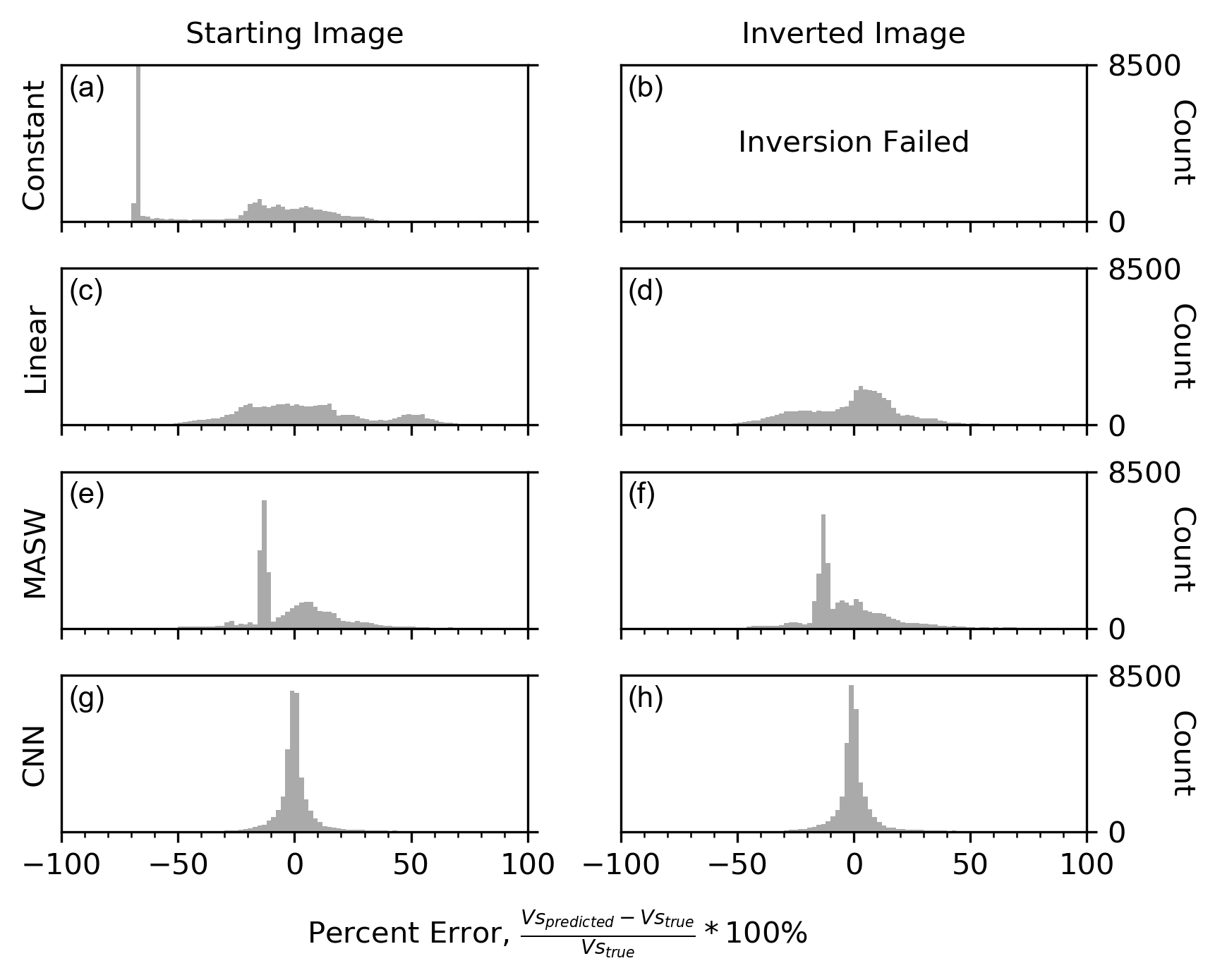}
	\caption{Distribution of the pixel-by-pixel percent error for the four starting models before and after full waveform inversion (FWI) for a \textbf{two-layered example} from the testing set. Panels (a), (c), (e), and (g) show the percent error for the constant, linear, multichannel analysis of surface waves (MASW), and convolutional neural network (CNN) starting images, adjacent to that of their inverted images after FWI (i.e., panels (b), (d), (f), and (h), respectively). Note that the distribution of percent error could not be calculated for the constant starting model, panel (b), as the inversion failed.}
	\label{fig:9}
\end{figure}

To quantitatively compare the starting and inverted Vs images relative to the true Vs image, the pixel-by-pixel Vs\textsubscript{residual} is shown in Figure \ref{fig:8}. Examining Figure \ref{fig:8}, we observe that in general all of the starting images, with the notable exception of the CNN, tend to overestimate the surface velocity (i.e., positive Vs\textsubscript{residual}) and underestimate the rock velocity (i.e., negative Vs\textsubscript{residual}). Furthermore, we observe that the constant starting image qualitatively resides the furthest from the true model, as it is primarily composed of large Vs\textsubscript{residual} (i.e., dark colors) and quantitatively has the largest MAPE = 39\%. Recall that it is this large difference between the constant starting model and the true model that is believed to have resulted in the failure of the FWI optimization. Comparing the other starting and inverted image pairs row-by-row, we observe that the largest inversion update occurred for the linear image, where improvements in the upper 1 -- 10 m are immediately apparent. This change is reflected in the decrease of its MAPE from 22\% to 16\%. The changes to the MASW and CNN starting images after FWI are almost un-notable in terms of both the spatial variability in the Vs\textsubscript{residual} values and the overall MAPE values, despite the MASW having a significantly larger MAPE than the CNN. The absence of any significant changes between the starting and inverted images will be discussed in greater detail when we compare the seismic waveforms in the following paragraph. Note that the thin region of Vs\textsubscript{residual}=0, which appears to delineate the bedrock interface in the panels of Figure \ref{fig:8}, is purely an artifact of how Vs\textsubscript{residual} is calculated and does not indicate the successful resolution of the boundary. For confirmation of this fact, refer back to Figure \ref{fig:7} and observe that the bedrock interface has not been clearly resolved. In summary, Figure \ref{fig:8} serves to highlight the good performance of the CNN relative to other approaches for generating FWI starting models, even though FWI does not significantly improve upon the MASW and CNN starting models. This finding is further demonstrated in Figure \ref{fig:9}, which shows the distributions of the pixel-by-pixel percent errors which are examined to detect bias. Importantly, the CNN approach is shown to not only produce a starting model with lower Vs\textsubscript{residual} and MAPE values than any of the other alternative starting models [compare panel (g) with panels (a), (c), and (e)], but it is also shown to outperform even those models obtained from FWI based on inferior starting models [compare panel (g) with panels (b), (d), and (f)]. These results clearly indicate the CNN approach alone (i.e., even without FWI) was able to produce a better representation of the true subsurface image (lower errors without bias) than was obtained using inferior starting models with the aid of FWI. Reasons for this somewhat surprising finding are discussed below.

\begin{figure}[!t]
	\centering
	\includegraphics[width=0.9\textwidth]{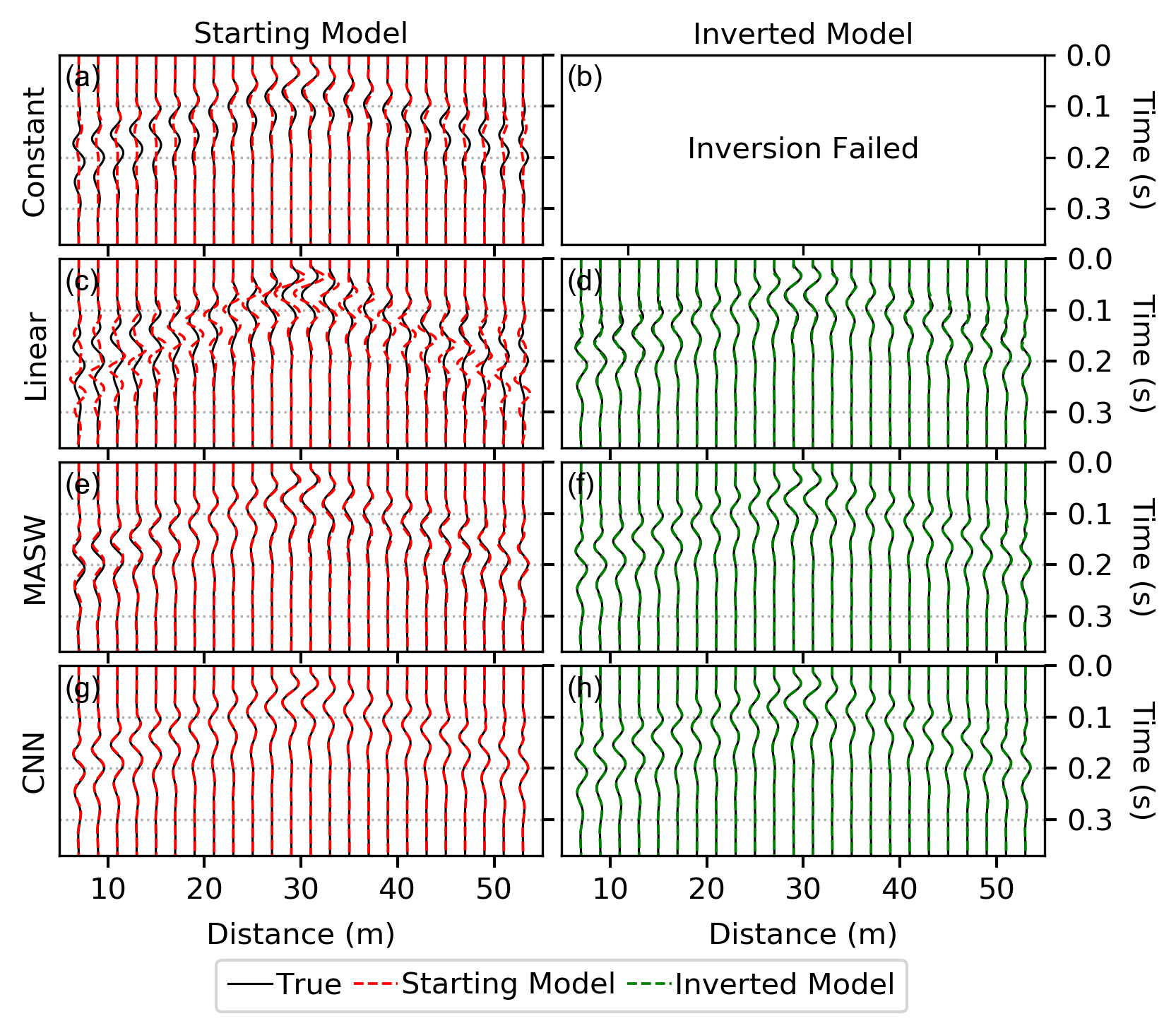}
	\caption{Qualitative comparison between synthetic waveforms before [panels (a), (c), (e), and (g)] and after [panels (b), (d), (f), and (h)] full waveform inversion with the true solution for a \textbf{two-layered model} from the testing set. The inversion of the constant starting model failed to produce results, therefore no comparison is made for the inverted model. Note the use of the term model (as opposed to image) is to remind the reader that while the waveforms presented are predominantly sensitive to the shear wave velocity image they are also affected, although to a lesser extent, by the other model parameters (compression wave velocity and mass density).}
	\label{fig:10}
\end{figure}

The waveforms from the central shot location (distance of 30 m) before and after FWI for the four starting models are compared in Figure \ref{fig:10}. The left column [i.e., panels (a), (c), (e), and (g)] compares the waveforms of the true model (solid black) and the corresponding starting model (dashed red). A quantitative comparison shows that as we progress to increasingly complex starting models [i.e., down the column from panel (a) to (g)] the fit to the true waveforms improves. A close examination of the constant starting model's waveforms in panel (a) shows that the first arrivals are generally well captured, however, the later shear and Rayleigh wave arrivals are not well captured, and at the further offsets, nearly an entire cycle of the surface wave is missing from the starting model's waveforms. For the linear starting model in panel (c), we observe that while the amplitude of the waveforms of the true model are not generally well captured, their trends are generally well captured, constituting an improvement from those resulting from the constant starting model. The MASW starting model in panel (e) is shown to capture the waveforms very well, with only minimal differences being observed at the farther distances and at times later in the record. Finally, the waveforms from the CNN starting model in panel (g) are shown to capture the true waveforms remarkably well across all distances and times. When comparing the waveforms of the inverted model for the three successful FWI inversions [i.e., panels (d), (f), and (h)], we observe the largest change for the linear starting model, some minimal changes for the MASW staring model, and essentially no change for the CNN starting model. This of course echoes the observations made with respect to Figures \ref{fig:7} and \ref{fig:8}, where the linear Vs starting image changed the most, followed by the MASW Vs starting image, and essentially no change being observed for the CNN Vs starting image after FWI. However, Figure \ref{fig:10} now provides insight into why this behavior should be expected; namely, that since the waveforms from the MASW and CNN starting models already fit the true waveforms quite well, large changes to these models during FWI were unnecessary to obtain low waveform misfit values. Furthermore, while the seismic wavefield from the MASW and CNN starting models match the seismic wavefield from the true model quite well [refer to Figure \ref{fig:10} panels (f) and (h)]; the Vs images from which they are derived are quite different [refer to Figure \ref{fig:7} panels (g) and (i)], highlighting the inherent non-uniqueness of the FWI inverse problem and emphasizing the influence of the starting model on the accuracy of the FWI results.

\begin{figure}[!t]
	\centering
	\includegraphics[width=0.6\textwidth]{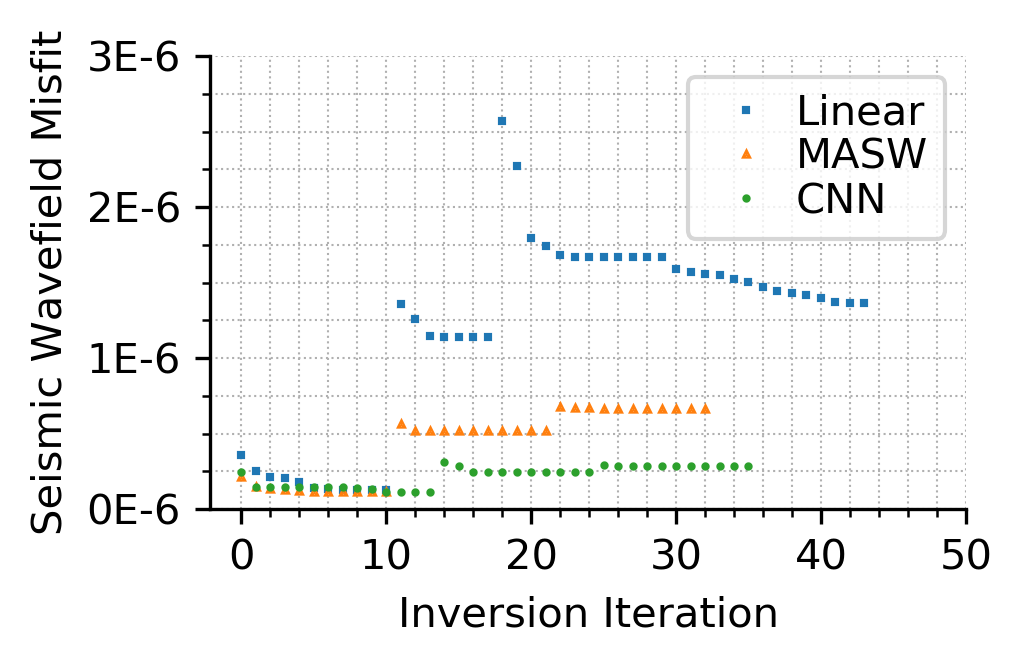}
	\caption{Seismic wavefield misfit as a function of inversion iteration for the three successful starting models for a \textbf{two-layered example} from the testing set. Note that the two discontinuous jumps to values of higher seismic wavefield misfit observed for all three starting models are the result of the inversion's transition between the three stages of the multi-scale inversion workflow, which included subsequent overlapping frequency ranges of 0 - 7 Hz, 0 - 13 Hz, and full signal.}
	\label{fig:11}
\end{figure}

To provide a quantitative comparison of the average waveform misfit across all shot locations, and to show the progression of the inversion process, the seismic wavefield misfit for each FWI iteration is presented in Figure \ref{fig:11}. Note the two discontinuous jumps to higher seismic wavefield misfits observed for all three starting models are the result of the inversion's transition between the three stages of the multi-scale inversion workflow. At the conclusion of the first inversion stage (near inversion iteration 10 in Figure \ref{fig:11}), all three starting models converge to nearly identical values of seismic wavefield misfit, indicating that the low frequency (0-7 Hz) components of the true model are well captured by each of the starting models. However, in the second and third inversion stages clear and significant differences in seismic wavefield misfit develop among the three starting models, with the CNN starting model ultimately achieving the lowest seismic wavefield misfit by a considerable margin. It is interesting to note that the seismic wavefield misfits of the MASW and CNN starting models change very little across the second and third inversion stage, whereas the linear model changes quite significantly. Yet, despite these significant updates, the linear model maintains the largest seismic wavefield misfit of the three alternatives (nearly two times larger than the MASW starting model's and nearly five times larger than the CNN starting model's seismic wavefield misfit), thereby making it the least desirable.

In this first example application of using a CNN to develop a starting model for FWI, we observe that the starting model has a significant effect on the FWI results, as the inverted model tends to retain characteristics from the starting model from which it was derived. So, in general, the better informed the starting model the more likely the inversion will converge to a reasonable approximation of the true model. More specifically, it is expected that starting models which are derived from either all (i.e., CNN) or part (i.e., MASW) of the measured wavefield will outperform those which are not (i.e., constant and linear), even under the most favorable conditions considered here (i.e., closest constant and linear Vs image to the true solution). In addition, we observe a clear example of the non-uniqueness of the FWI problem by showing that models which are quite different (refer to Figure \ref{fig:7}e, g, and i) can result in waveforms that appear quite similar (refer to Figure \ref{fig:10}d, f and, h) and fit the waveforms of the true seismic image quite well.

\section{Comparison of the CNN-Derived Starting Model with other Starting Models in FWI: Three-Layered Example}

\begin{figure}[!t]
	\centering
	\includegraphics[width=0.9\textwidth]{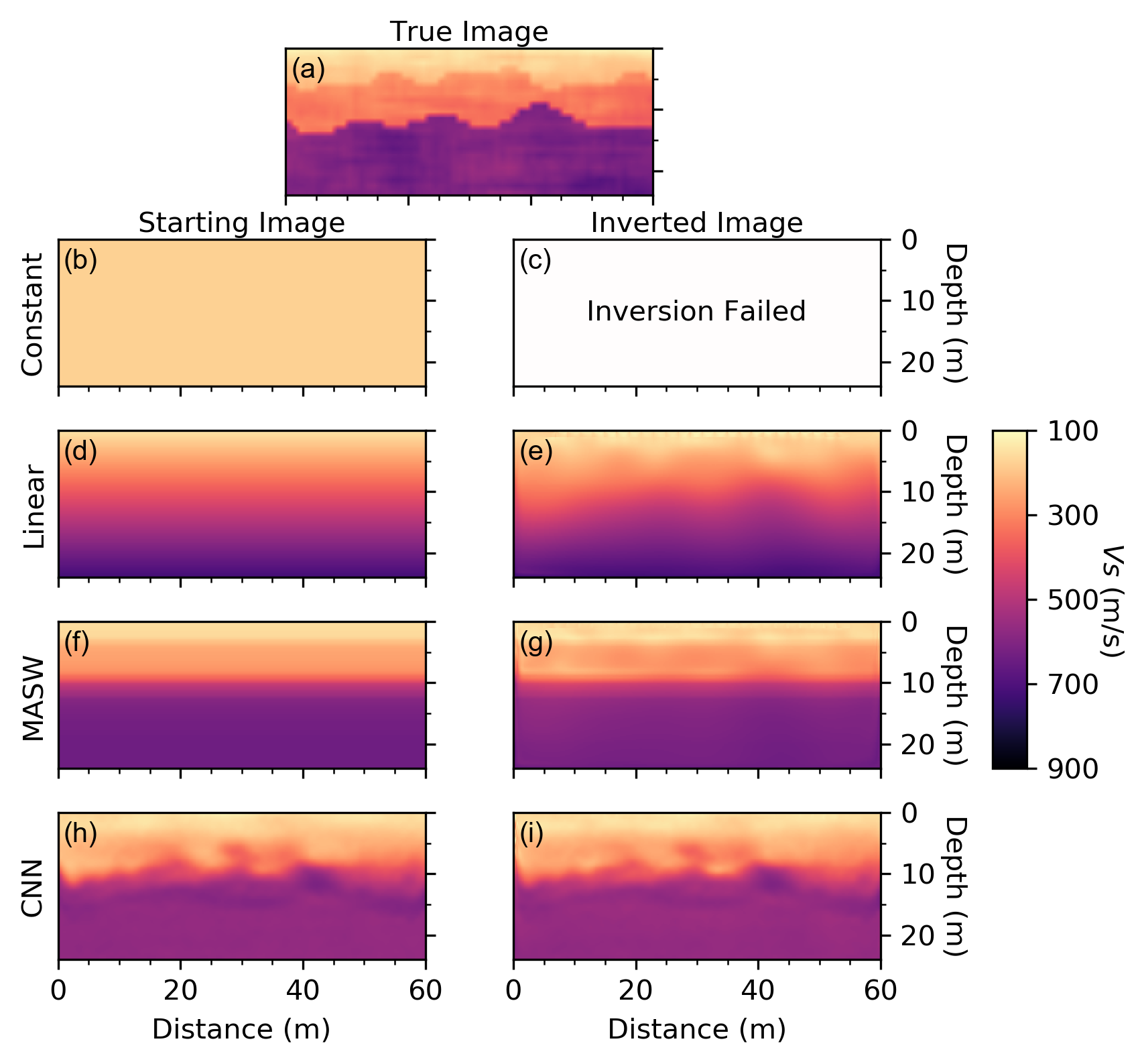}
	\caption{Performance of the convolutional neural network (CNN) in comparison to other typical starting models for the \textbf{three-layered example}.  Panels (b), (d), (f), and (h) show the constant, linear, multichannel analysis of surface waves (MASW), and convolutional neural network (CNN) shear wave velocity (Vs) starting image, adjacent to their inverted Vs images after full waveform inversion (FWI) (i.e., panels (c), (e), (g), and (i), respectively). For reference the true Vs image is shown in panel (a). Note that no inverted image is available for the constant starting model, panel (c), as the inversion failed.}
	\label{fig:12}
\end{figure}

To assess the ability of the CNN to generalize to other subsurface images which are dissimilar from the population upon which it was trained (i.e., two-layered systems), we now consider a system with three undulating layers. These layers were selected to consist of approximately 5 m of soil (Vs $\approx$100 m/s), overlying approximately 5 m of intermediate geomaterials (Vs $\approx$300 m/s), overlying weathered rock (Vs $\approx$600 m/s). The true seismic image is shown in Figure \ref{fig:12}a. To compare the performance of the CNN-derived starting model, we will again consider starting models as in the previous section: constant, linear, and MASW-derived. The Vs images of the four starting models are presented in panels (b), (d), (f), and (h) of Figure \ref{fig:12}. With the exception of adjusting the comparison depth from the upper 10 m to the upper 5 m for the constant model, the constant, linear, MASW, and CNN starting models were developed in a manner identical to that described in the previous section using the new three-layered true model and its associated waveforms. The FWI of the four starting models followed the same procedures outlined in the previous example. Again, the constant starting model failed to produce any inversion results (refer to Figure \ref{fig:12}c), as the waveform optimization failed due to instability prior to the conclusion of the first iteration of the first stage. Multiple attempts were made with different inversion parameters, however, ultimately no FWI solution was possible with a constant starting model. Of the three successfully inverted models, the linear starting model again undergoes the most significant update during FWI, followed by the MASW starting model, and finally the CNN starting model. We qualitatively observe that the CNN-derived starting model and subsequent inverted model are not as accurate as the previous example, however, this is to be expected, as the three-layered system resides outside the CNN model's two-layered training space. Nonetheless, visual inspection of the CNN Vs starting image (Figure \ref{fig:12}h) and its accompanying Vs inverted image (Figure \ref{fig:12}i) shows that they are most similar to the true Vs image (Figure \ref{fig:12}a). These results are encouraging, because by using the CNN to develop a starting model for a three-layered model we were knowingly pushing the limits of the CNN, which was trained using only two-layered models. Importantly, while three-layered models were not included as part of the present CNN's training set, they could be easily incorporated in the future through additional training. This additional training, while quite easy to do, was not performed for the present study, as the purpose of this example is to illustrate the CNN's ability to adapt and generalize to systems dissimilar from the population on which it was trained.

\begin{figure}[!t]
	\centering
	\includegraphics[width=0.9\textwidth]{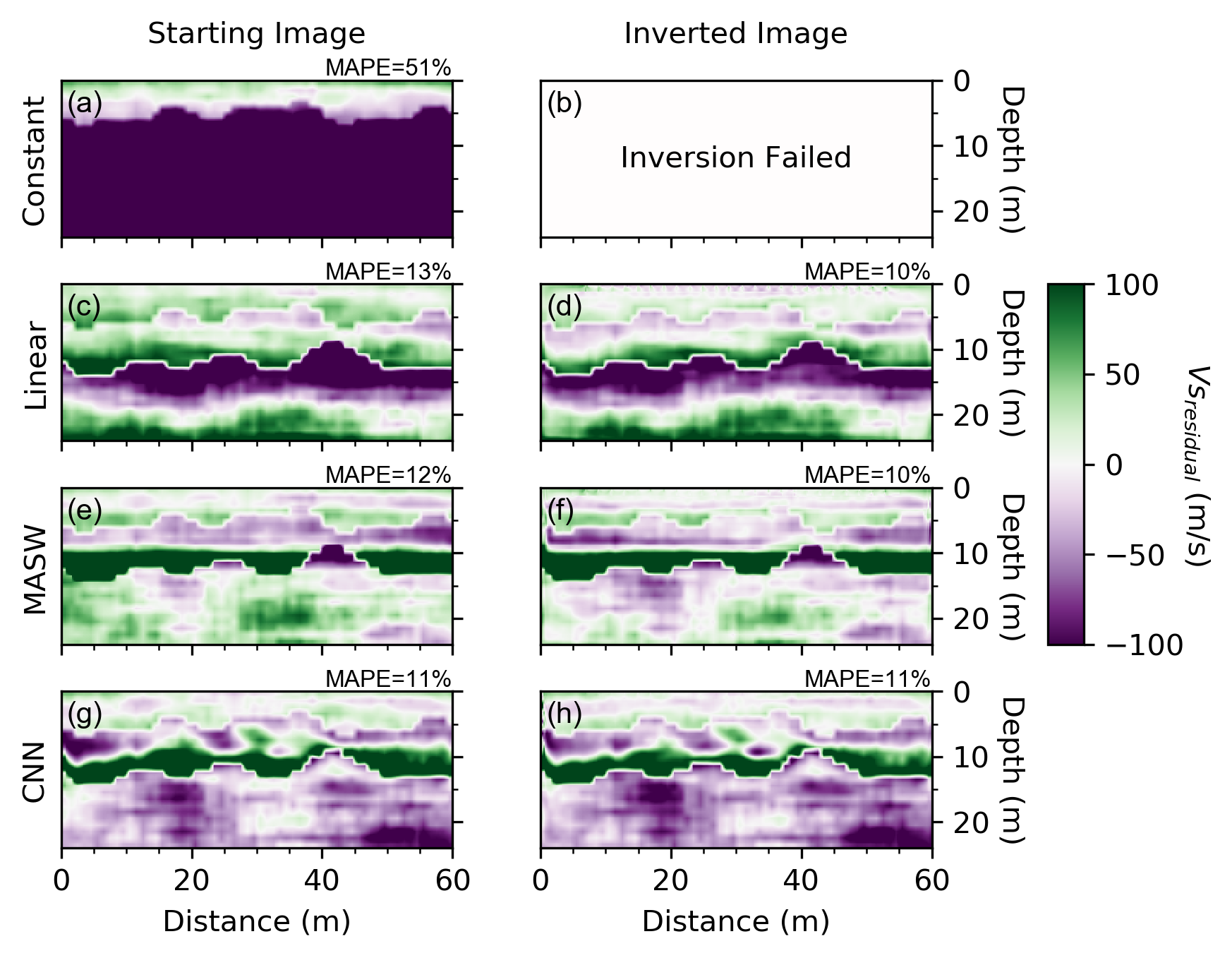}
	\caption{The residual shear wave velocity (Vs\textsubscript{residual}) for the four starting models before and after full waveform inversion (FWI) for the \textbf{three-layered example}. Panels (a), (c), (e), and (g) show Vs\textsubscript{residual} for the constant, linear, multichannel analysis of surface waves (MASW), and convolutional neural network (CNN)  starting images, adjacent to that of the inverted images after FWI (i.e., panels (b), (d), (f), and (h), respectively). The mean absolute percent error (MAPE) of each image is presented in the upper right of each panel. Note that the thin region of Vs\textsubscript{residual}=0 which appears to delineate the bedrock interface is purely an artifact of how Vs\textsubscript{residual} is calculated and does not indicate the successful resolution of the boundary. Note that Vs\textsubscript{residual} could not be calculated for the constant starting model, panel (b), as the inversion failed.}
	\label{fig:13}
\end{figure}

\begin{figure}[!t]
	\centering
	\includegraphics[width=0.9\textwidth]{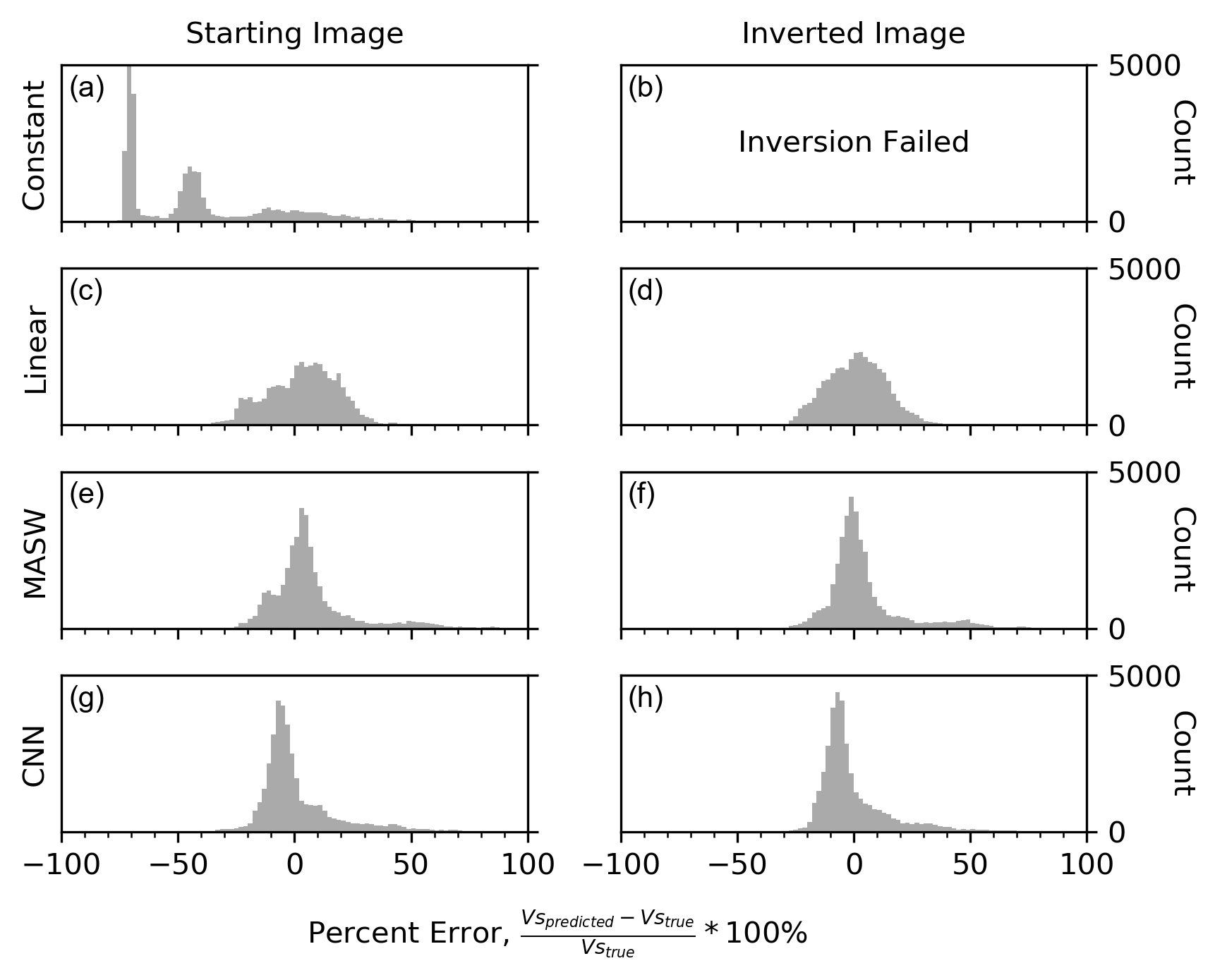}
	\caption{Distribution of the pixel-by-pixel percent error for the four starting models before and after full waveform inversion (FWI) for the \textbf{three-layered example}. Panels (a), (c), (e), and (g) show the percent error for the constant, linear, multichannel analysis of surface waves (MASW), and convolutional neural network (CNN) starting images, adjacent to that of their inverted images after FWI (i.e., panels (b), (d), (f), and (h), respectively). Note that the distribution of percent error could not be calculated for the constant starting model, panel (b), as the inversion failed.}
	\label{fig:14}
\end{figure}

Figure \ref{fig:13} presents the pixel-by-pixel Vs\textsubscript{residual} of the four starting models before and after FWI. Several important points can be drawn from Figure \ref{fig:13}. First, the constant Vs starting image (Figure \ref{fig:13}a) is, with the exception of the top 5 m, primarily composed of large Vs\textsubscript{residual}, as evidenced by its dark colors and large MAPE of 51\%. These large Vs\textsubscript{residual} (recall Figure \ref{fig:8}a) are likely to blame for the model's failure to update during FWI. Second, when comparing the successful starting models (linear, MASW, and CNN, panels (c), (e), and (g), respectively) it is less clear than the previous example [refer to Figure \ref{fig:7} panels (c), (e), and (g)] which of the three is the best. While all three have similar MAPE values before (between 11\% and 13\%), and after (between 10\% and 11\%) inversion [refer to Figure \ref{fig:12} panels (d), (f), and (h), respectively], as noted previously, it could easily be argued that based on a visual comparison the CNN's Vs image after FWI yields the most realistic representation of the true Vs image. Figure \ref{fig:14}, which presents the pixel-by-pixel percent error, provides some support for this qualitative assessment, as the CNN's error after FWI [i.e.,  panel (h)], while slightly biased towards lower Vs, is more compact than that of the linear and MASW images [i.e.,  panels (d) and (f), respectively]. Of course, while quantitative numbers are desirable, caution should be exercised to avoid over-interpreting the informative ability of any single quantitative measure. For example, the distribution of percent error of the MASW Vs image after FWI [Figure \ref{fig:14} (f)] appears to indicate it performed the best, with minimal bias and relatively small percent errors. Yet, visual examination of the FWI results obtained from the MASW Vs image after FWI [Figure \ref{fig:12} (g)] indicates all of the geologic interfaces as horizontal, in stark opposition to the true Vs image [Figure \ref{fig:12} (a)], which has a number of clear undulations that are clearly better represented by the CNN starting model both prior to and after FWI [Figure \ref{fig:12} panels (h) and (i)]. 

\begin{figure}[!t]
	\centering
	\includegraphics[width=0.9\textwidth]{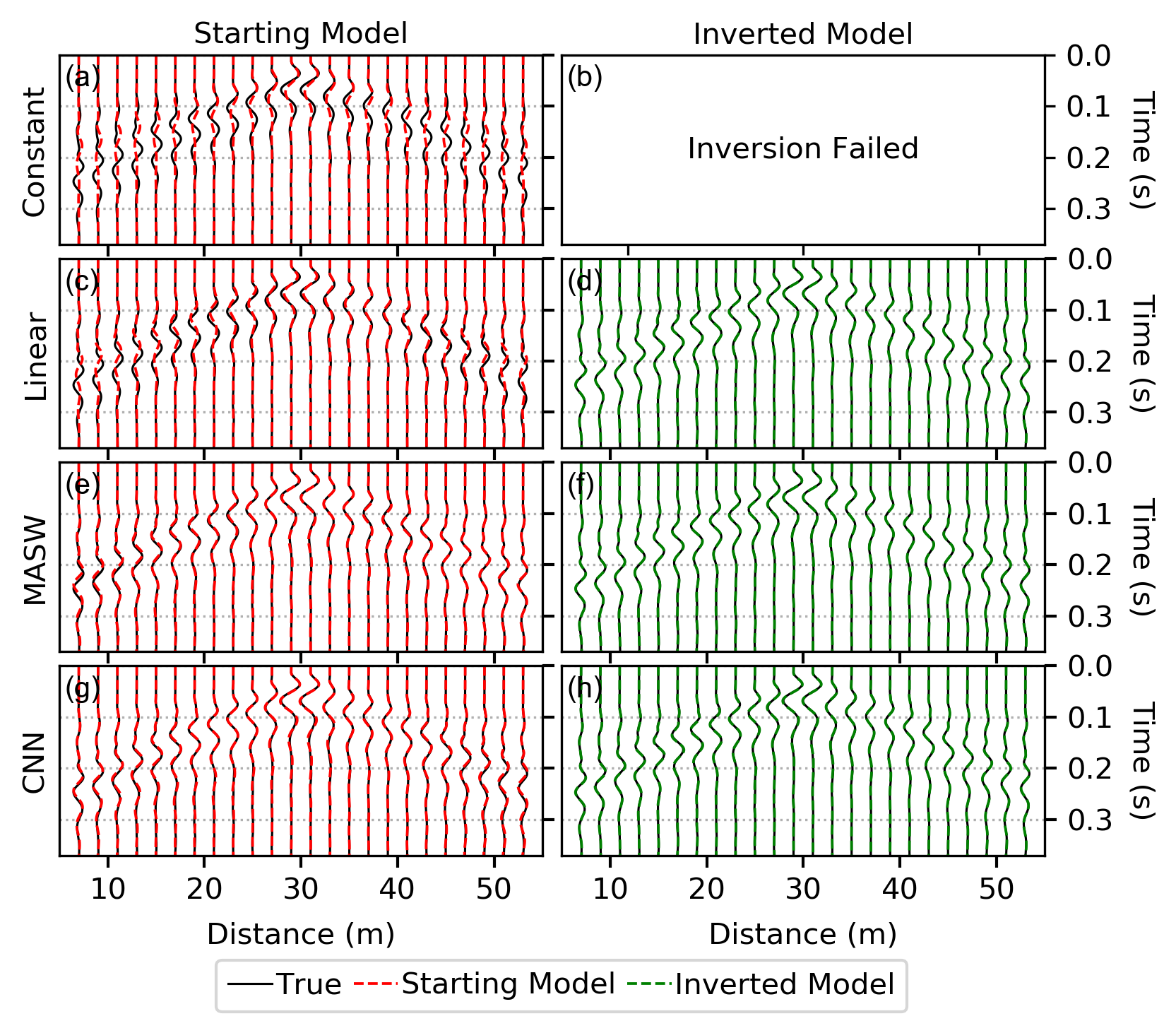}
	\caption{Qualitative comparison between synthetic waveforms before [panels (a), (c), (e), and (g)] and after [panels (b), (d), (f), and (h)] full waveform inversion with the true solution for the \textbf{three-layered example}. The inversion of the constant starting model failed to produce results, therefore no comparison is made for the inverted model. Note the use of the term model (as opposed to image) is to remind the reader that while the waveforms presented are predominantly sensitive to the shear wave velocity image they are also affected, although to a lesser extent, by the other model parameters (compression wave velocity and mass density).}
	\label{fig:15}
\end{figure}

\begin{figure}[!t]
	\centering
	\includegraphics[width=0.6\textwidth]{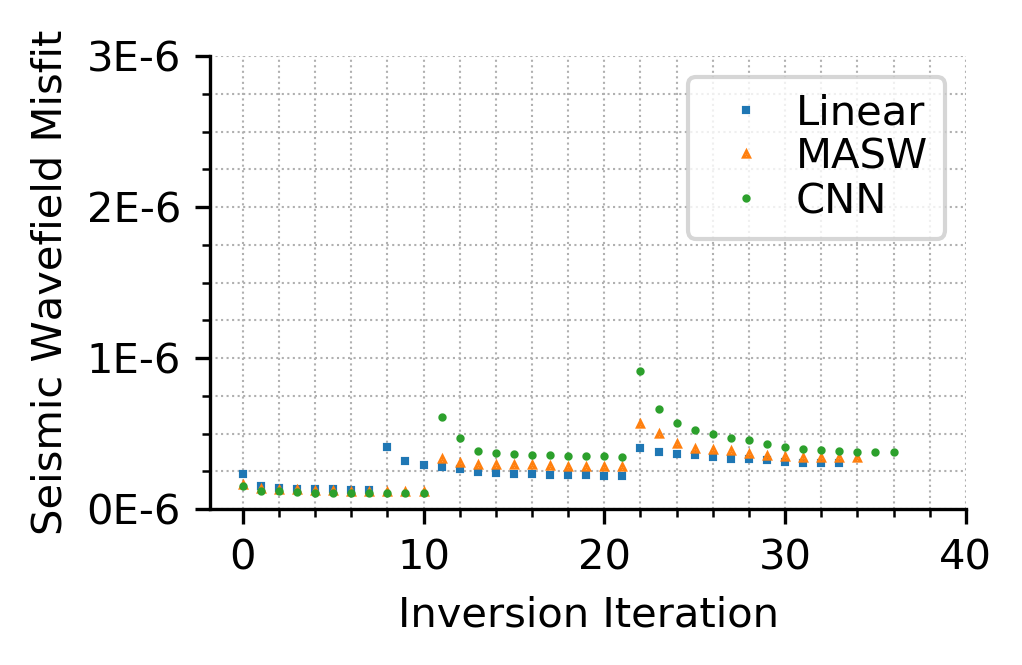}
	\caption{Seismic wavefield misfit as a function of inversion iteration for the three successful starting models for the \textbf{three-layered example}. Note that the two discontinuous jumps to values of higher seismic wavefield misfit observed for all three starting models are the result of the inversion's transition between the three stages of the multi-scale inversion workflow, which included subsequent overlapping frequency ranges of 0 - 7 Hz, 0 - 13 Hz, and full signal.}
	\label{fig:16}
\end{figure}

As with the previous two-layered example, the waveforms for the three-layered example, shown in Figure \ref{fig:15}, help to explain the FWI model update behavior observed in Figures \ref{fig:12} and \ref{fig:13}. First, in Figure \ref{fig:15}a, we observe that the constant starting model fails to approximate the majority of the true model's waveforms, with large differences being observed over the majority of the time record. The presence of these large differences highlights the inadequacy of a constant starting model for this example. Second, the agreement between the waveforms of the various starting models [i.e., panels (c), (e), and (g)] and the true model again appear to increase according to each approach's ability to utilize the measured seismic wavefield, with the linear model fitting the waveforms the worst and the CNN fitting the waveforms the best. Interestingly, however, when comparing the waveforms after FWI, panels (d), (f), and (h) show that all three starting models qualitatively produce waveform fits of equally good quality, despite the observations with regard to Figure \ref{fig:12}(e), (g), and (i) that the models from which they were derived are quite different. To show the similarity between the waveforms in a more quantitative manner, and to highlight the effect of the inversion process, the wavefield misfit is presented in Figure \ref{fig:16} for each iteration of FWI. Unlike the previous example, we observe that all three starting models produce FWI results (in terms of their seismic wavefield misfit) of essentially equal quality, with the linear and MASW final inverted models slightly outperforming the CNN final inverted model in terms of waveform misfit. Furthermore, we observe very small changes in misfit across the three stages of the multi-scale workflow in comparison to the significant changes in misfit observed in the previous example (recall Figure \ref{fig:11}). The similarity of the misfits between the three starting models at the end of FWI further highlights the non-uniqueness inherent in inverse problems, such as FWI, as models that appear quite different (recall Figure \ref{fig:12}) can produce waveforms which are qualitatively (recall Figure \ref{fig:15}) and quantitatively (recall Figure \ref{fig:16}) quite similar, making it necessary to communicate the potential for such non-uniqueness in practice, and work towards developing procedures for its quantification.

Finally, it is important to provide a brief discussion of the CNN derived model in relation to the other two successful starting models for the three-layered case and provide insight into why the CNN derived model did not perform as well as in the two-layered case. First, by comparing Figures \ref{fig:11} and \ref{fig:16} we observe that in terms of the wavefield misfit at the end of FWI, the CNN for the three-layered example performed only slightly worse than it did for the two-layered example, indicating that the CNN was able to learn features/representations from the two-layered seismic wavefield-image pairs it was trained on that generalize to other seismic images and was still capable of producing an adequate (although imperfect) starting model. Second, from Figures \ref{fig:11} and \ref{fig:16}, we observe a large change in the performance of the linear staring model, which went from performing the worst in terms of the seismic wavefield misfit to the best. This is explained by recognizing that since the true model (refer to Figure \ref{fig:12}a) contains three layers rather than two, by definition it includes weaker impedance contrasts between subsurface layers and becomes inherently more linear. Therefore, it should be expected that the performance of the linear model will improve as the true subsurface model is more linear, with less abrupt impedance contrasts. Third, examining the true three-layered model, the undulations of the interfaces are more smooth and less frequent laterally, and can therefore be better approximated assuming a 1D starting model, such as those provided by the linear and MASW starting models. Therefore, it should be expected that these starting models will perform better than they did previously. Finally, since the CNN was only trained on two-layered systems, we should not expect it to perform well when analyzing a more complicated three-layered system. However, that is not to say the CNN-derived starting model is poor, as it is able to produce a seismic wavefield misfit after inversion despite these complicating factors that is essentially equivalent to that of the other two approaches. Furthermore, by visual inspection, the CNN-derived starting model and its post-FWI counterpart could easily be argued to produce subsurface models that are most representative of the true model (refer to Figure \ref{fig:12}). This example serves to highlight that CNNs show significant promise as a tool for developing starting models for FWI, but before they can be deployed in practice additional work is required to train the network using a large variety of subsurface models to ensure sufficient generality to be applied routinely in practice.

\section{Conclusions}

This paper assesses the efficacy of using CNNs to develop starting models for 2D FWI. The CNN developed in this work focuses on the classic near-surface imaging problem of identifying the location of a two-layer undulating soil-bedrock interface. After training on 200,000 pairs of synthetic seismic waveform-image pairs, the CNN showed excellent performance, being capable of generating subsurface seismic image predictions, based solely on the recorded seismic wavefield, for a 100 model testing set with a mean absolute percent error (MAPE) of 6\%. To compare the predictive ability of the CNN-derived starting models to other typical approaches for generating FWI starting models, a single model from the testing set was inverted using the CNN-derived starting model and three other common starting models: constant, linear, and MASW-derived. The CNN was able to produce a starting model which resided closer to the true solution in terms of Vs than the other approaches and produce smaller waveform misfits before and after FWI. These results indicate that CNNs can be used to develop starting models for FWI and that those models can outperform other common alternatives. To assess the ability of the CNN to generalize to subsurface models which are dissimilar to the ones upon which it was trained, a second example which contains two material interfaces (i.e., a three-layered system) was inverted using the CNN's prediction in comparison to the same three common starting model approaches. While the predictive ability of the CNN was slightly reduced, it was still able to achieve seismic image and wavefield misfits comparable to, or better than, the other starting models, despite the example's clear disadvantage to the CNN. Both examples highlight the need to develop accurate starting models for FWI, as the choice of the starting model is shown to strongly influence the quality of the final inverted model. Furthermore, these examples illustrate that the use of different starting models can result in inverted models which may appear quite different but fit the experimental seismic wavefield data equally well, emphasizing that in order to capture the true solution an initial model that is a good approximation of the subsurface conditions is required. This study demonstrates the need for developing accurate starting models for FWI and that CNNs show great promise as a potential solution.

\section{Acknowledgements}

The forward solution and full waveform inversion (FWI) algorithm used in this study are from the open-source software DENISE \citep{kohn_time_2011, kohn_influence_2012}. The construction of the seismic wavefield-image pairs and performance of traditional full waveform inversion (FWI) used the Texas Advanced Computing Center's (TACC's) cluster Stampede2 using an allocation provided through DesignSafe-CI \citep{rathje_designsafe_2017}. The training of the convolutional neural network (CNN) used TACC's machine and deep learning resource Maverick2. The multichannel analysis of surface wave (MASW) data was processed using the open-source Python package \textit{swprocess} \citep{vantassel_jpvantasselswprocess_2021} and inverted using the Dinver module \citep{wathelet_surface-wave_2004} of the open-source software Geopsy \citep{wathelet_geopsy_2020}. The figures in this paper were created using Matplotlib 3.1.2 \citep{hunter_matplotlib_2007} and Inkscape 0.92.4. This work was supported by the U.S. National Science Foundation grant CMMI-1931162. However, any opinions, findings, and conclusions or recommendations expressed in this material are those of the authors and do not necessarily reflect the views of the National Science Foundation.

\bibliographystyle{plainnat}
\bibliography{fwi_cnn}

\end{document}